\documentclass[11pt,twoside]{article}

\usepackage{amsmath}
\usepackage{amsfonts}
\usepackage{amssymb}
\usepackage{latexsym}
\usepackage{graphicx}
\newcommand{\opn}[1]{\operatorname{#1}}
\newcommand{\veps}{\varepsilon}
\newcommand{\real}{\opn{Re}}
\newcommand{\imag}{\opn{Im}}
\newcommand{\phm}{\phantom{-}}
\newcommand{\mlc}[1]{\multicolumn{1}{r}{#1}}
\newcommand{\dder}[2]{\frac{d #1}{d #2}}
\newcommand{\der}[2]{\frac{\partial #1}{\partial #2}}
\newcommand{\varder}[2]{\frac{\delta #1}{\delta #2}}
\newcommand{\e}[1]{{(#1)}}
\newcommand{\mc}[1]{\mathcal{#1}}
\newcommand{\jt}{\textstyle}

\newcommand{\js}{\scriptstyle}

\usepackage{epsfig}

\usepackage{theorem}
\theoremstyle{plain}
\theoremheaderfont{\scshape}
\newtheorem{theorem}{Theorem}

\title{Computation of Time-Periodic Solutions of the \\
  Benjamin-Ono Equation}

\date{July 2, 2009}

\author{
David M. Ambrose
\thanks{Department of Mathematical Sciences, Clemson
  University, Clemson, SC 29634.
  Current address: Department of Mathematics, Drexel University,
  Philadelphia, PA 19104 ({\tt ambrose@math.drexel.edu}).
  This work was supported in part by the National Science Foundation
  through grant DMS-0926378.}
\and
Jon Wilkening
\thanks{Department of Mathematics and Lawrence Berkeley National
  Laboratory, University of California, Berkeley, CA 94720 ({\tt
    wilken@math.berkeley.edu}).  This work was supported in part by the
  Director, Office of Science, Computational and Technology Research,
  U.S. Department of Energy under Contract No. DE-AC02-05CH11231.}
}

\begin{document}

\maketitle

\begin{abstract}
  We present a spectrally accurate numerical method for finding
  non-trivial time-periodic solutions of non-linear partial
  differential equations.  The method is based on minimizing a
  functional (of the initial condition and the period) that is
  positive unless the solution is periodic, in which case it is zero.
  We solve an adjoint PDE to compute the gradient of this functional
  with respect to the initial condition.  We include additional terms
  in the functional to specify the free parameters, which, in the case
  of the Benjamin-Ono equation, are the mean, a spatial phase, a
  temporal phase and the real part of one of the Fourier modes at
  $t=0$.

  We use our method to study global paths of non-trivial time-periodic
  solutions connecting stationary and traveling waves of the
  Benjamin-Ono equation.  As a starting guess for each path, we
  compute periodic solutions of the linearized problem by solving an
  infinite dimensional eigenvalue problem in closed form.  We then use
  our numerical method to continue these solutions beyond the realm of
  linear theory until another traveling wave is reached.
  By experimentation with data fitting, we
  identify the analytical form of the solutions on the path connecting
  the one-hump stationary solution to the two-hump traveling wave.  We
  then derive exact formulas for these solutions by explicitly solving
  the system of ODEs governing the evolution of solitons
  using the ansatz suggested by the numerical simulations.
\end{abstract}

\vspace*{5pt}

\vspace*{5pt}

\pagestyle{myheadings}
\markboth{David M. Ambrose and Jon Wilkening}{Time-Periodic Solutions
of the Benjamin-Ono Equation}

\section{Introduction}

A fundamental problem in the theory of ordinary and partial
differential equations is to determine whether the equation
possesses time-periodic solutions.  Famous examples of ordinary
differential equations with periodic solutions include the Brusselator
\cite{field:burger,govaerts,strogatz:chaos} and the three-body problem
\cite{arenstorf,hairer1}.  In partial differential equations,
time-periodic solutions can be ``trivial'' stationary or traveling
waves, or can be genuinely time-periodic.  Such a problem can be
studied in either the forced or unforced context.  Forced problems
include an external force in the PDE that is time-periodic;
solutions with the same period are then sought.  In the unforced
problem, the period is one of the unknowns.  In this work, we present
a numerical method for finding genuinely time-periodic solutions of
the unforced Benjamin-Ono equation with periodic boundary conditions.
These solutions have many remarkable properties, which we will
describe.

Our work is motivated by the calculations of Hou, Lowengrub, and
Shelley for the vortex sheet with surface tension \cite{HLS1,HLS2},
and by the analysis of Plotnikov, Toland and Iooss
\cite{tolandPlotnikov, tolandPlotnikovIooss} for the water wave.  Hou,
Lowengrub, and Shelley developed an efficient numerical method to
solve the initial value problem for the vortex sheet with surface
tension.  They performed calculations for a variety of initial
conditions and values of the surface tension parameter, and found many
situations in which the solutions appear to be close to time-periodic.
They did not, however, try to measure the deviation from
time-periodicity or attempt to vary the initial conditions to reduce
this deviation.  Plotnikov, Toland, and Iooss have proved the
existence of time-periodic water waves, without surface tension, in
the case of either finite or infinite depth.  This is proved using a
version of the Nash-Moser implicit function theorem.  Their work
includes no computation of the water waves.  We aim to get a firmer
handle on these solutions with an explicit calculation.  To this end,
in the present work, we develop a general numerical method for finding
time-periodic solutions of nonlinear systems of partial differential
equations and eventually plan to use this method for the vortex sheet
and water wave problems.  In fact, during the review process of
the present work, we have succeeded in computing several families of
time-periodic solutions of the vortex sheet with surface tension
\cite{ambrose:wilkening:vtx}.

The Benjamin-Ono equation, developed in
\cite{benjamin:67,davisAcrivos,ono}, is a model equation for the
evolution of waves on deep water.  It is a widely-studied dispersive
equation, and much is known about solutions.  It would be impossible
to mention all results on Benjamin-Ono, but we mention, for example,
that weak solutions exist for $u_{0}\in L^{2}$ \cite{saut,
  ginibreVelo}, and that the solution exists for all time if $u_{0}\in
H^{1}$ \cite{tao}.  We chose the Benjamin-Ono equation as a first
application for our numerical method because it is much less expensive
to evolve than the vortex sheet or water wave, yet has many features
in common with them, such as non-locality (due to the Hilbert
transform in the former case and the Birkhoff-Rott integral in the
latter two cases.)  In our numerical simulations, we find that the
Benjamin-Ono equation has a rich family of non-trivial time-periodic
solutions that act as rungs in a ladder connecting traveling waves
with different speeds and wavelengths by creating or annihilating
oscillatory humps that grow or shrink in amplitude until they become
part of the stationary or traveling wave on the other side of the
rung.  The dynamics of these non-trivial solutions are often very
interesting, sometimes resembling a low amplitude traveling wave
superimposed on a larger carrier signal, and other times looking like
a collection of interacting solitons that pass through each other or
bounce off each other, depending on their relative amplitudes.

By fitting our numerical data, we have determined that all the
solutions we have computed are special cases of the multi-phase
solutions studied by Case \cite{case:mero}, Satsuma and Ishimori
\cite{satsuma:ishimori:79}, Dobrokhotov and Krichever \cite{dobro:91},
and Matsuno \cite{matsuno:04}, but with special initial conditions
(that yield periodic orbits) and a modified mean (to change their
speeds and allow bifurcations to occur between different levels of the
hierarchy of multi-phase solutions).  We did not take advantage of
this structure when we developed our numerical method; hence, our
approach can also be used for non-integrable problems.  We also note
that bifurcation within the family of multi-phase solutions has not
previously been discussed in the literature, nor has the remarkable
dynamics of the Fourier coefficients of these solutions beyond the
original derivation by Benjamin \cite{benjamin:67} of the form of the
traveling waves for this equation.

We are aware of very few works on the existence of time-periodic
solutions for water wave model equations.  Crannell has
demonstrated \cite{crannell} the existence of periodic, non-traveling,
weak solutions of the Boussinesq equation using a generalization of
the mountain pass lemma of Rabinowitz.  Chen and Iooss have proved
existence of time-periodic solutions in a two-way Boussinesq-type
water wave model \cite{chenIooss}.  As in \cite{tolandPlotnikov} and
\cite{tolandPlotnikovIooss}, there is no computation of the solution
in either of these studies.  Cabral and Rosa have recently discovered a
period-doubling cascade of periodic solutions for a damped and forced
version of the Korteweg--de Vries equation \cite{cabral:rosa}.  They
use a Fourier pseudospectral method for the spatial discretization and
a first order semi-implicit scheme in time.  To find periodic
solutions, they use a secant method on a numerical Poincar\'{e} map.
Whereas our approach is based on minimizing a functional that measures
deviation from periodicity, they rely on the stability of the orbit to
converge to a periodic solution. They stop when they find a solution
that returns to within one percent of its initial state, whereas we
resolve our periodic solutions to 13-15 digits of accuracy, which
allows us to study the analytic form of the solutions.

Water waves aside, many authors have investigated time-periodic
solutions of other partial differential equations both numerically and
analytically.  For instance, Smiley proves existence of time-periodic
solutions of a nonlinear wave equation on an unbounded domain
\cite{smileyReine}; he also develops a numerical method for the same
problem \cite{smileyNumerical}.  On a finite domain, Brezis uses
duality principles to prove the existence of periodic solutions of
nonlinear vibrating strings in both the forced and unforced setting;
see \cite{brezis}.  Mawhin has written a survey article on periodic
solutions of semilinear wave equations \cite{mawhin}, which includes
many references.  Pao has developed a numerical method for the
solution of time-periodic parabolic boundary-value problems
\cite{pao}.  Pao gives various iterative schemes, but unlike the
present work, these are not based on variational principles or the
dual system.  Brown et.~al.~\cite{kevrekidis} have used the
  popular software package AUTO~\cite{doedel91a} to study Hopf
  bifurcation and loss of stability for traveling waves of the
  regularized Kuramoto-Sivashinsky equation.  They observe modulated
  traveling waves similar to many of the solutions we have found for
  Benjamin-Ono.  And of course, time-periodic solutions of systems of
ordinary differential equations have also been widely studied,
e.g.~in \cite{rabinowitz78,rabinowitz82,zehnder,duistermaat}.

 The most popular methods for the numerical solution of boundary
  value problems governed by ODEs, which include finding time-periodic
  solutions as a special case, are orthogonal collocation methods
  \cite{doedel91b} and shooting/multi-shooting methods
  \cite{stoer:bulirsch}.  These methods can also be used for PDEs via
  the method of lines, i.e.~by discretizing space first to obtain an
  ODE.  Orthogonal collocation methods (such as
  implemented in AUTO) require a nonlinear system of equations to be
  solved involving all the degrees of freedom at every timestep.  In
  many of our simulations, the solution at every timestep cannot even
  be stored simultaneously in memory; hence, it would be impossible to
  solve a system of equations for all these variables.  For PDEs,
  shooting methods are also very expensive as the Jacobian of the
  functional measuring errors in the boundary conditions must be
  computed with respect to variation of the $N$ ``unknown'' initial
  conditions, which is $N$ times more expensive than computing the
  functional itself.

 Lust and Roose \cite{lust:roose:92} propose an interesting
  solution to this problem in which a shooting method is used for some
  of the degrees of freedom while iteration on a Poincar\'{e} map is
  used for the other degrees of freedom.  They aim to exploit the fact
  that many high dimensional systems actually exhibit low-dimensional
  dynamical behavior.  In the present work, we offer an alternative
  strategy based on the following idea: while the Jacobian of the
  residual measuring error in each of the boundary conditions is
  expensive to compute, the gradient of the sum of squares of this
  residual can be computed by solving a single adjoint PDE.  Although
  the full Jacobian gives more information, it is $N$ times more
  expensive to compute. We find that it is more efficient to use a
  quasi-Newton method with only the gradient information than it is to
  use a full Newton method with the Jacobian.  This approach is not
  restricted to finding time-periodic solutions; it will work for any
  boundary value problem.  Also, although we have not tried it, our
  method can be easily adapted to incorporate the idea behind
  multi-shooting methods, namely that by introducing ``interior''
  initial conditions and boundary conditions, we can avoid a great
  deal of the ill conditioning due to nearby trajectories diverging
  exponentially in time.

The closest numerical method to our own that we have found is due to
Bristeau, Glowinski and P\'eriaux \cite{glowinski1}, who developed a
least squares shooting method for numerical computation of
time-periodic solutions of linear dynamical systems with applications
in scattering phenomena in two and three dimensions; see also
\cite{glowinski2}.  These authors employ methods of control theory to
compute variational derivatives, and although they only apply their
methods to linear problems, they mention that their techniques will
also work on non-linear problems.  Our method can be considered an
extension of their approach that focuses on the difficulties that
arise due to non-linearity.  In particular, we replace their conjugate
gradient solver with a black-box minimization algorithm, (the BFGS
method \cite{nocedal}), and include an additional penalty function to
prescribe the values of the free parameters that describe the manifold
of non-trivial time-periodic solutions.  Without this penalty
function, the basic method is only found to produce constant solutions
and traveling waves.

This paper is organized as follows: In Section \ref{stationary}, we
discuss spatially periodic stationary and traveling solutions of the
Benjamin-Ono equation, the bifurcations from constant solutions to
traveling waves, and the pole dynamics of meromorphic solutions.
In Section \ref{linearTheory}, we investigate
time-periodic solutions of the linearized Benjamin-Ono equation; this
is the linearization about the stationary solutions discussed
previously.  To analyze the linearized problem, we compute
(numerically) the spectrum and eigenfunctions of the relevant linear
operator and deduce their analytic form by trial and error; the
resulting formulas can be verified rigorously (but we omit details).

In Section \ref{method}, we describe our numerical method, which
involves minimizing a non-negative functional that is zero if and only
if the solution is periodic.  We solve an adjoint PDE to compute the
variational derivative of this functional with respect to perturbation
of the initial condition and use the BFGS minimization algorithm to
minimize the functional.  The Benjamin-Ono and adjoint equations are
solved with a pseudo-spectral collocation method using a fourth order,
semi-implicit Runge-Kutta scheme.  We use a penalty function to rule
out constant solutions and traveling waves, and to prescribe the free
parameters of the manifold of non-trivial solutions.  We then vary a
bifurcation parameter to study the global properties of
these non-trivial solutions.  In the present work, we apply this
method only to the Benjamin-Ono equation, but we are confident that
this method is applicable to virtually any system of partial
differential equations that possesses time-periodic solutions.

In Section \ref{sec:nontrivial}, we use our method to study the global
behavior of non-trivial time-periodic solutions far beyond the realm
of validity of the linearization about stationary and traveling waves.
We will follow one such path to discover that the one-hump stationary
solution is connected to the two-hump traveling wave by a path of
non-trivial time periodic solutions.  In Section~\ref{sec:exact}, we
re-formulate the ODE governing the evolution of poles to reveal an
exact formula for the solutions on the path studied numerically in
Section~\ref{sec:nontrivial}. Thus, unexpectedly, we have proved that
non-trivial time-periodic solutions bifurcate from stationary
solutions by exhibiting a family of them explicitly.  In a follow-up
paper \cite{benj:ono2}, we will classify all bifurcations from
traveling waves, study the paths of non-trivial solutions connecting
several of them, propose a conjecture explaining how they all fit
together, and describe their analytic form to the extent that we are
able.  We end with a few concluding remarks in
Section~\ref{conclusion}.

\section{Stationary, Traveling and Soliton Solutions}
\label{stationary}

We consider the spatially periodic Benjamin-Ono equation with the
following sign convention:
\begin{equation} \label{eqn:BO}
  u_{t}=Hu_{xx}-uu_{x}.
\end{equation}
Of course, the operator $H$ is the Hilbert transform.  Recall that the
symbol of $H$ is $\hat{H}(k)=-i\operatorname{sgn}(k).$
  Many authors include a factor of two in the convection term and
  place a minus sign in front of $H$.  The former change causes
  solutions to be scaled by a factor of $1/2$ while the latter change
  has no effect as $H$ is defined with the opposite sign with this
  convention.

\begin{figure}
\begin{center}
\includegraphics[width=.7\linewidth]{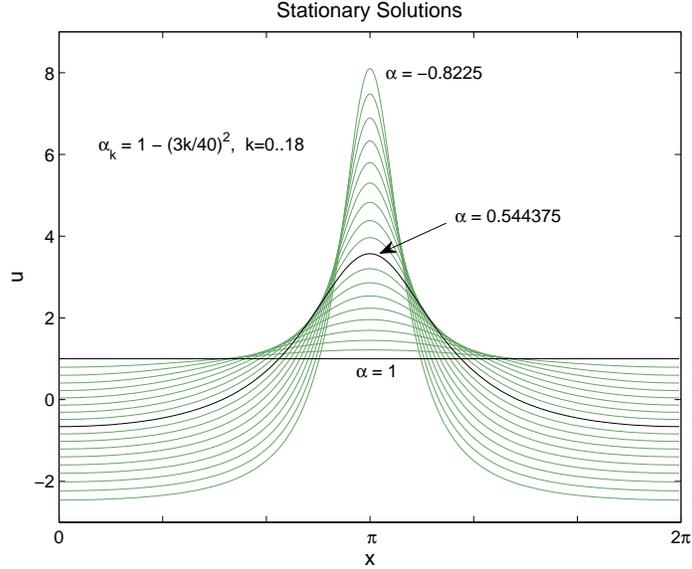}
\end{center}
\caption{Stationary solutions of the Benjamin-Ono equation.}
\label{fig:stationary}
\end{figure}

This equation possesses a
two-parameter family of stationary solutions, namely
\begin{equation} \label{eqn:stationary}
  u(x)=\frac{1-3\beta^{2}}{1-\beta^{2}} +
  \frac{4\beta[\cos(x-\theta) - \beta]}
  {1 + \beta^{2}-2\beta\cos(x-\theta)},
  \qquad (-1<\beta<1,\;\theta\in\mathbb{R}).
\end{equation}
These solutions have mean $\alpha$, related to $\beta$ via
\begin{equation}\label{eqn:alphabeta}
\alpha(\beta) = \frac{1 - 3|\beta|^2}{1 - |\beta|^2}, \qquad
|\beta|^{2} = \frac{1 - \alpha}{3 - \alpha}.
\end{equation}
Changing the sign of $\beta$ is equivalent to the phase shift
$\theta\rightarrow\theta-\pi$.  It is convenient to complexify
$\beta$ and define $u_\beta$ to be the mean-zero part of
(\ref{eqn:stationary}) with $\beta\rightarrow|\beta|$,
$\theta\rightarrow \opn{arg}\bar{\beta}$:
\begin{equation}
  \text{stationary solution} = \alpha(\beta) + u_\beta(x), \qquad
  \beta = |\beta|e^{-i\theta} \in\Delta=\{z\in\mathbb{C}:|z|<1\}.
\end{equation}
Note that the subscript $\beta$ does not indicate a derivative here.
Several stationary solutions with $\beta$ real and negative
are shown in Figure~\ref{fig:stationary}.  The Fourier representation
of $u_\beta$ is simply 
\begin{equation}
\hat{u}_{\beta,k}=\left\{\begin{array}{ll}2\bar{\beta}^{|k|}, & k<0 \\
 0, & k=0 \\ 2\beta^{k}, & k>0\end{array}\right\},
\end{equation}
where $\bar\beta$ is the complex conjugate of $\beta$.
These functions $u_\beta(x)$ are the building blocks
for the meromorphic solutions discussed below.

Note that the constant solution $u\equiv\alpha_0$ is also a stationary
solution, as are the re-scaled solutions
$$u_{N,\beta}(x) = N\alpha(\beta) + N u_{\beta}(Nx),
\qquad (\beta\in\Delta,\; N=1,2,3,\dots),$$
which have mean $\alpha_0=N\alpha(\beta)$.  If we restrict 
attention to stationary solutions with even symmetry (i.e.~with
$\beta$ real), we find that there is a pitchfork bifurcation at each
positive integer (using the mean as a bifurcation parameter).
As $\alpha_0$ changes from $N^+$ to $N^-$, the constant solution
splits, yielding two additional ($N$-hump stationary) solutions,
namely $u_{N,\beta}(x)$ with $\beta=0^\pm$.  The pitchfork would be
obtained by plotting the real part of the $N$th Fourier mode versus
the mean, where we observe that the Fourier representation of
$u=u_{N,\beta}$ (for any $\beta\in\Delta$) is given by
\begin{equation} \label{eqn:uhat:k}
  \hat{u}_k = \begin{cases}
    N\alpha(\beta), & k=0, \\
    2N\beta^{k/N}, & k\in N\mathbb{Z}, \; k>0, \\
    2N\bar{\beta}^{|k|/N}, & k\in N\mathbb{Z}, \; k<0, \\
    0 & \text{otherwise.}
  \end{cases}
\end{equation}
If we do not restrict attention to even solutions, the phase shift
$\theta$ acts as a second parameter connecting the two outer branches
of the pitchfork into a two-dimensional, bowl-shaped sheet (plotting
the real and imaginary parts of the $N$th Fourier mode versus the
mean).

In the bifurcation problem just described, we varied the mean
$\alpha_0$ and found bifurcations from constant solutions to
stationary solutions at the positive integers.  The remainder of this
paper deals with bifurcation from these stationary solutions to
non-trivial time-periodic solutions and their global continuation
beyond the realm of linear theory.  Rather than varying the mean, we
will hold $\alpha_0\in\mathbb{R}$ constant and use another quantity
(such as the period $T$ or the real part of one of the Fourier modes
of $u$ at $t=0$) as the bifurcation parameter.  As a first step, let
us consider bifurcation from constant solutions to traveling waves
holding $\alpha_0\in\mathbb{R}$ constant and varying $T$.

All traveling wave solutions of the Benjamin-Ono equation can be found
by applying a simple transformation to a stationary solution, and vice
versa.  Indeed, if $u(x,t)$ is any solution of (\ref{eqn:BO}), then
\begin{equation} \label{eqn:add:trav}
  U(x,t)=u(x-ct,t)+c
\end{equation}
is also a solution; thus, adding a constant $c$
to a stationary solution causes it to travel to the right with speed
$c$.  We can parametrize these $N$-hump traveling waves by their mean
$\alpha_0\in\mathbb{R}$ and decay/phase parameter $\beta\in\Delta$:
\begin{equation} \label{eqn:trav:soln:def}
  u_{\alpha_0,N,\beta}(x,t) = u_{N,\beta}(x-ct)+c, \qquad
  \big(c = \alpha_0 - N\alpha(\beta)\big).
\end{equation}
If we express the period $T=2\pi/(N|c|)$ in terms of $\beta$ and solve
for $\beta$, we find that we can bifurcate from any constant solution
$u\equiv\alpha_0$ to an $N$-hump traveling solution with the same
mean.  If $\alpha_0<N$, a pitchfork from the constant solution occurs
at $T_0=2\pi/[N(N-\alpha_0)]$; as we increase $T$ from $T_0$ to
$\infty$, $\alpha=[2\pi/(NT)+\alpha_0]/N$ decreases from 1 to
$\alpha_0/N$ and $|\beta|$ varies from 0 to
$\sqrt{(1-\alpha_0/N)/(3-\alpha_0/N)}$.  Similarly, if
$\alpha_0>N$, a pitchfork occurs at $T_0=2\pi/[N(\alpha_0-N)]$; as we
\emph{decrease} $T$ from $T_0$ to 0, $\alpha=[\alpha_0-2\pi/(NT)]/N$
decreases from 1 to $-\infty$, and $|\beta|$ varies from 0 to $1$.
And if $\alpha_0=N$, the situation is qualitatively similar to the
latter case, but the pitchfork occurs at $T_0=\infty$, i.e.~all three
solutions (with $\beta$ real) exist for any period $T>0$.

We remark that if the traveling waves described above have zero mean,
they are a special case of the meromorphic $N$-particle ``periodic
soliton'' solutions described in \cite{case:BO:N:sol}, namely
\begin{equation*}
  u(x,t) = 2\real\left\{\sum_{l=1}^N \frac{2}{e^{i[x+t-x_l(t)]}-1}
  \right\},
\end{equation*}
where $\imag\{x_l(0)\}>0$ and the $x_l(t)$ satisfy the
system of differential equations
\begin{equation} \label{eqn:soliton:ode}
  \frac{dx_l}{dt} = \sum_{\parbox{.3in}{$\js m=1\\[-8pt]m\ne l$}}^N
  \frac{2}{e^{-i(x_m-x_l)}-1}
  + \sum_{m=1}^N \frac{2}{e^{-i(x_l-\bar{x}_m)}-1}, \qquad
  (1\le l\le N).
\end{equation}
In our notation, we write
\begin{equation*}
  x_l(t) = \overline{i\log\beta_l(t)} =
  \theta_l(t) - i\log|\beta_l(t)|, \qquad (\beta_l=|\beta_l|e^{-i\theta_l}=
  e^{-i\bar{x}_l})
\end{equation*}
and generalize to the case that the mean $\alpha_0$ can be non-zero.
We find from (\ref{eqn:soliton:ode}) and (\ref{eqn:add:trav})
that
\begin{equation}\label{eqn:mero:soln}
  u(x,t)=\alpha_0+\sum_{l=1}^N u_{\beta_l(t)}(x)
\end{equation}
is a solution of (\ref{eqn:BO}) if the variables $\beta_l\in
\Delta$ satisfy
\begin{equation} \label{eqn:beta:ode}
  \dot{\beta}_l = \sum_{\parbox{.3in}{$\js m=1\\[-8pt]m\ne l$}}^N
  \frac{-2i\beta_l^2}{\beta_l-\beta_m} +
  \sum_{m=1}^N \frac{2i\beta_l^2}{\beta_l - \bar{\beta}_m^{-1}}
  + i(2N-1-\alpha_0)\beta_l, \qquad (1\le l\le N).
\end{equation}
The $N$-hump traveling wave then has the representation
\begin{equation*}
  u_{\alpha_0,N,\beta}(x,t) = \alpha_0 + \sum_{l=1}^N u_{\beta_l(t)}(x),
  \qquad \beta_l(t) = \sqrt[N]{\beta} e^{-ict}, \qquad
  c = \alpha_0 - N\alpha(\beta),
\end{equation*}
where each $\beta_l$ is assigned a distinct $N$th root of $\beta$.  As
we are interested in developing numerical methods that generalize to
more complicated systems such as the vortex sheet with surface tension
and the water wave, we do not exploit the existence of meromorphic
solutions in our numerical method; however, the non-trivial time-periodic
solutions we find do turn out to be of this form; see
Section~\ref{sec:exact}.

\section{Linear Theory}\label{linearTheory}

We formulate the problem of finding time-periodic solutions of the
Benjamin-Ono equation as that of finding an initial condition $u_0$
and period $T$ such that $F(u_0,T)=0$, where
$F:H^1\times\mathbb{R}\rightarrow H^1$ is given by
\begin{equation} \label{eqn:F:def}
  F(u_0,T) = u(\cdot,T)-u_{0}, \qquad u_t = Hu_{xx} - uu_x, \quad
  u(\cdot,0)=u_0.
\end{equation}
Clearly, stationary solutions are periodic with any period $T$.
In this section, we linearize $F$ about these stationary solutions
and use solutions of the linearized problem as initial search
directions to find time-periodic solutions of the nonlinear problem.
Bifurcation from traveling waves can be reduced to this case by adding
an appropriate constant and requiring that the period of the
perturbation coincide with the period of the traveling wave (although
there may be a phase shift involved as well).  We present a detailed
analysis of the traveling case in \cite{benj:ono2}.

\subsection{Linearization About Stationary Solutions} \label{sec:lin:stat}

Let $u=u_{N,\beta}$ be an arbitrary $N$-hump stationary solution.  If
$u(x) + v(x,t)$ is to satisfy (\ref{eqn:BO}) to first order in $v$,
then $v$ should satisfy
\begin{equation} \label{eqn:BO:lin}
  v_t = Hv_{xx} - (uv)_x.
\end{equation}
(The exact solution satisfies $v_t=Hv_{xx}-(uv)_x-vv_x$.)  Equation
(\ref{eqn:BO:lin}) can be written
\begin{equation} \label{eqn:vt:BA}
  v_{t}=iBAv,
\end{equation}
where the (unbounded, self-adjoint) operators $A$ and $B$ on $H^1$
are defined as
\begin{equation} \label{eqn:AB:def}
  A=H\partial_{x} - u,\qquad B=\frac{1}{i}\partial_{x}.
\end{equation}
To solve (\ref{eqn:vt:BA}), we are interested in the eigenvalue problem
\begin{equation} \label{eqn:BA:eval:problem}
  BAz = \omega z,
\end{equation}
so that if $BA$ has a complete set of eigenvectors, the
general solution of (\ref{eqn:vt:BA}) will be a superposition
of functions of the form
$$v(x,t) = \real\{C z(x) e^{i\omega t}\}, \qquad C\in\mathbb{C}.$$
Of course, the eigenvalues of a composition of Hermitian operators
need not be real, but for $A$ and $B$ in (\ref{eqn:AB:def}), we can
compute all the eigenvalues explicitly, and they are indeed real.  We
do this numerically (which surprisingly leads us to formulas we can
check analytically) by truncating the Fourier representations of $A$
and $B$ and computing the eigenvalues of the matrix $\hat{B}\hat{A}$.
More precisely, we choose a cutoff frequency $K$ (e.g.~$K=240$) and
define the $(2K-1)\times(2K-1)$ matrices
\begin{equation}
  \hat{A}_{kl} = |k|\delta_{kl} - \hat{u}_{k-l} =
  |k|\delta_{kl} - \overline{\hat{u}_{l-k}} , \qquad
  \hat{B}_{kl} = k\delta_{kl}, \qquad (-K < k,l < K),
\end{equation}
where $\hat{u}_k$ was given in (\ref{eqn:uhat:k}) and $\delta_{kl}=1$
if $k=l$ and 0 otherwise.  By carefully studying the eigenvalues for
different values of $N$ and $\beta=-\sqrt{(1-\alpha)/(3-\alpha)}$
with $\alpha<1$, we determined that
\begin{equation} \label{eqn:omega:Nn}
  \omega_{N,n} = \begin{cases}
    -\omega_{N,-n} & n<0, \\
    \phm 0 & n=0, \\
    (n)(N-n)
      & 1\le n\le N-1, \\
      (n+1-N)\big(n+1+N(1-\alpha)\big)\quad &
      n\ge N.
\end{cases}
\end{equation}
With this numbering, the first $N-1$ non-zero eigenvalues
are independent of $\alpha$:
\begin{equation}\label{eqn:omega:Nn:table}
\begin{array}{c|c}
   & n \\ \hline
 N & \omega_{N,n}
\end{array} \quad = \quad
\begin{array}{c|r|r|r|r|r|rr}
    & \mlc{1} & \mlc{2} & \mlc{3} & \mlc{4} & \mlc{5} & \mlc{6} & \cdots \\ \hline
  1 & \mlc{*} & \mlc{*} & \mlc{*} & \mlc{*} & \mlc{*} & \mlc{*} & \cdots \\ \cline{2-2}
  2 &      1  & \mlc{*} & \mlc{*} & \mlc{*} & \mlc{*} & \mlc{*} & \cdots \\ \cline{3-3}
  3 & \mlc{2} &      2  & \mlc{*} & \mlc{*} & \mlc{*} & \mlc{*} & \cdots \\ \cline{4-4}
  4 & \mlc{3} & \mlc{4} &      3  & \mlc{*} & \mlc{*} & \mlc{*} & \cdots \\ \cline{5-5}
  5 & \mlc{4} & \mlc{6} & \mlc{6} &      4  & \mlc{*} & \mlc{*} & \cdots \\ \cline{6-6}
  6 & \mlc{5} & \mlc{8} & \mlc{9} & \mlc{8} &      5  & \mlc{*} & \cdots
\end{array}
\end{equation}
Note that $\omega_{N,N}=(2-\alpha)N+1\ge N+1$ and $\omega_{N,n}$ is
strictly increasing in $n$ for $n\ge N$, but $\omega_{N,N}$ could be
less than $\omega_{N,\lfloor N/2 \rfloor}$ when $N\ge6$ (and some of
the eigenvalues can coalesce, increasing their multiplicity).
Nevertheless, the ordering of the eigenvalues in (\ref{eqn:omega:Nn})
is more convenient than the monotonic ordering due to the fact that a
pathway of non-trivial solutions connecting an $N$-hump traveling wave
to an $N'$-hump traveling wave with $N<N'$ seems to involve
$\omega_{N,n}$ and $\omega_{N',n'}$ with $n\ge N$ and $n'<N'$
satisfying $N'=n+1$ and $n'=N'-N$ (see \cite{benj:ono2}.)  These
global reconnection formulas would be much more complicated if the
eigenvalues were ordered monotonically.

The zero eigenvalue $\omega_{N,0}=0$ has geometric multiplicity two
and algebraic multiplicity three.  The fact that the dimension of the
kernel is independent of $\alpha$ indicates that there are no special
values of the mean $N\alpha$ at which these $N$-hump stationary
solutions bifurcate to more complicated stationary solutions.
The two eigenfunctions in the kernel of $BA$ are
\begin{equation} \label{eqn:kernel:BA}
  z_{N,0}^\e{1,0}(x) =
  -u_x(x) = \der{}{\theta}\bigg\vert_{\theta=0} u_{N,\beta}(x-\theta),
  \qquad z_{N,0}^\e2(x) = \der{}{|\beta|} u_{N,\beta}(x),
\end{equation}
which correspond to translating the stationary solution by a phase or
decreasing its mean, $N\alpha=N(1-3|\beta|^2)/(1-|\beta|^2)$.  There
is also a Jordan chain \cite{Achain} of length two associated with
$z_{N,0}^\e{1,0}(x)$, namely
\begin{equation}
  z_{N,0}^\e{1,1}(x) = i, \qquad \left(BA z_{N,0}^\e{1,1} =
    z_{N,0}^\e{1,0}\right).
\end{equation}
The corresponding solution of (\ref{eqn:vt:BA}) is
$$v(x,t) = -i z_{N,0}^\e{1,1}(x) + tz_{N,0}^\e{1,0}(x) = 1 - t u_x(x) =
\der{}{\veps}\bigg\vert_{\veps=0} [u(x-\veps t)+\veps],$$
i.e.~this linear growth mode arises due to the fact that adding a
constant to a stationary solution causes it to travel.  The multiple
eigenvalues $\omega_{N,n}=\omega_{N,N-n}$ with $1\le n\le N-1$ pose
a minor obstacle to obtaining explicit formulas for the eigenvectors.
We eventually realized that because the shift operator
\begin{equation} \label{eqn:S:def}
  S_\theta z(x) = z(x-\theta), \qquad
  \hat{S}_{\theta,kl} = e^{-ik\theta}\delta_{kl}, \qquad
  \theta=2\pi/N
\end{equation}
commutes with $BA$, the eigenspaces of $BA$ are invariant under the
action of $S_\theta$.  Thus we can impose the additional requirement
that if $z$ is an eigenvector of $BA$ corresponding to a multiple
eigenvalue, then $z$ should also satisfy
\begin{equation}\label{eqn:zk:additional}
\hat{z}_k\ne0 \quad \text{and} \quad \hat{z}_l\ne0 \quad
\Rightarrow \quad k-l\in N\mathbb{Z},
\end{equation}
i.e.~the non-zero Fourier coefficients are equally spaced with stride
length $N$.  Using this condition to make the eigenvectors unique up
to scaling, we were able to recognize the patterns that
emerge in the numerical eigenvectors (with the exception of the
coefficient $C$ and the $j=0$ case when $n\ge N$, which we determined
analytically):
\begin{align}
  \notag
  &\hat{z}_{N,n,k}\Big\vert_{k=n+jN} = \left\{\begin{array}{cc}
      \left(1+\frac{N(|j|-1)}{N-n}\right)\bar\beta^{|j|-1} & j<0 \\[5pt]
      C\left(1 + \frac{Nj}{n}\right)\beta^{j+1} & j\ge0
      \end{array}\right\},  \qquad
    \Bigg(\parbox{1.56in}{\begin{center}$1\le n\le N-1$ \\
          $C = \frac{-n N}{(N-n)\big[n+(N-n)|\beta|^2\big]}$
        \end{center}}\Bigg), \\[5pt]
\label{eqn:evec:formulas}
 &\hat{z}_{N,n,k}\Big\vert_{k=n-N+1+jN} = \left\{\begin{array}{cc}
    0 & j<0 \\[5pt]
    \frac{-\bar\beta}{(1-|\beta|^2)^2}\left[
      1 - \left(1-\frac{N}{n+1}\right)|\beta|^2\right] & j=0 \\[5pt]
    \left(1 + \frac{N(j-1)}{n+1}\right)\beta^{j-1} & j\ge1
  \end{array}\right\}, \qquad(n\ge N).
\end{align}
These formulas can be summed to obtain $z_{N,n}(x)$ as a rational
function of $e^{ix}$, but we prefer to work with the Fourier
coefficients.  Note that as $n\rightarrow\infty$ (holding $N$ fixed),
the index $k=n-N+1$ of the first non-zero Fourier mode increases to
infinity.  The eigenvectors corresponding to negative eigenvalues
$\omega_{N,-n}$ with $n\ge1$ satisfy
$z_{N,-n}(x)=\overline{z_{N,n}(x)}$, so the Fourier coefficients
appear in reverse order, conjugated: $\hat{z}_{N,-n,k} =
\overline{\hat{z}_{N,n,-k}}$.  When $\beta$ is real, the Fourier
coefficients are real and $z_{N,-n}(x)=z_{N,n}(-x)$.  We have verified
the formulas (\ref{eqn:omega:Nn}) and (\ref{eqn:evec:formulas})
analytically, and can also prove that the Fourier representation of
these eigenvectors (together with the associated vector corresponding
to the Jordan chain) form a Riesz basis for $\ell^2(\mathbb{Z})$;
hence, we have not missed any eigenvalues.

\subsection{Bifurcation from Stationary Solutions}\label{sec:bif:stat}

Now that we have solved the eigenvalue problem for $BA$, we can
compute the derivative of the operator $F$ in (\ref{eqn:F:def}) above.
We continue to assume that $u$ is an N-hump stationary solution so
that $DF=(D_1F,D_2F):H^1\times\mathbb{R}\rightarrow H^1$ satisfies
\begin{equation}
  \begin{aligned}
    D_1F(u,T)v_0 &= \der{}{\veps}\Big\vert_{\veps=0}F(u+\veps v_0,T) =
    v(\cdot,T)-v_0 = \left[e^{iBAT}-I\right]v_0, \\[2pt]
    D_2F(u,T)\tau &= \der{}{\veps}\Big\vert_{\veps=0}F(u,T+\veps\tau) = 0.
  \end{aligned}
\end{equation}
Note that $v_0\in\ker D_1F(u,T)$ iff the solution $v(x,t)$ of the
linearized problem is periodic with period $T$.  These 
solutions of the linearized problem serve as initial search directions
in which to find periodic solutions of the non-linear problem.
Since $D_2F=0$ in the stationary case, the nullspace
$\mc{N}$ of $DF=(D_1F,D_2F)$ is of the form
$\mc{N}=\mc{N}_1\times\mathbb{R}$ where $\mc{N}_1=\ker D_1F$.
A basis for $\mc{N}_1$ consists of the functions
\begin{equation} \label{eqn:v0:re:im}
  v_0(x) = \real\{z_{N,n}(x)\} \qquad \text{ and } \qquad
  v_0(x) = \imag\{z_{N,n}(x)\},
\end{equation}
where $n$ ranges over all integers such that
\begin{equation}\label{eqn:omega:condition}
  \omega_{N,n}T\in \, 2\pi\mathbb{Z}.
\end{equation}
Negative values of $n$ have already been accounted for in
(\ref{eqn:v0:re:im}) using $z_{N,-n}(x)=\overline{z_{N,n}(x)}$.
The $n=0$ case always yields two vectors in the kernel, namely those
in (\ref{eqn:kernel:BA}).  These directions do not cause bifurcations
as they lead to other stationary solutions in the two parameter
family.  Thus, the periods at which bifurcations are expected are
\begin{equation} \label{eqn:T:Nnm}
  T_{N,n,m} = \frac{2\pi m}{\omega_{N,n}}, \qquad (m,n\ge1).
\end{equation}
Note that this set is dense on the positive real line since
$\omega_{N,n}\rightarrow\infty$ as $n\rightarrow\infty$.  This leads
to a small divisor problem when trying to apply the Liapunov-Schmidt
reduction \cite{golubitsky,kielhofer} to rigorously analyze
bifurcations in the set of solutions of the equation $F(u_0,T)=0$.
For other problems, small divisors have been dealt with successfully
using Nash-Moser theory; see e.g.~\cite{nirenberg, tolandPlotnikov,
  tolandPlotnikovIooss}.  As most of the bifurcations responsible for
the small divisor problem involve high frequency perturbations that
are smoothed out by the numerical discretization, we have not found
small divisors to cause major difficulties in our ability to track
paths of time-periodic solutions.

In our numerical studies, we have found that each of the
  eigenfunctions $z_{N,n}(x)$ with $n\ge1$ gives a direction along
  which a sheet of non-trivial solutions bifurcates from the $N$-hump
  stationary solution.  More precisely, if we let a parameter $\veps
  \rightarrow0$, there appear to be time-periodic solutions of the
  nonlinear problem with period $T_\veps=T_{N,n,m}+O(\veps^2)$ that
  agree with the real and imaginary parts of $\veps z_{N,n}(x)$ at
  $t=0$ and $t=-T_\veps /(4m)$, respectively, to $O(\veps^2)$.  This
  is interesting because in cases that several (even infinitely many)
  of the eigenfunctions $z_{N,n}(x)$ lead to solutions of the
  linearized problem with the same period, most linear combinations of
  these solutions will not give a bifurcation direction.  In other
  words, the eigenfunctions $z_{N,n}(x)$ are already diagonalized with
  respect to the nonlinear effects of the bifurcation.
  For this to work out, it is essential that when $\omega_{N,n}$ is a
  multiple eigenvalue, $z_{N,n}(x)$ is chosen as in
  (\ref{eqn:zk:additional}) to simultaneously diagonalize the shift
  operator $S_{2\pi/N}$.

\begin{figure}[t]
\begin{center}
\includegraphics[width=.6\linewidth]{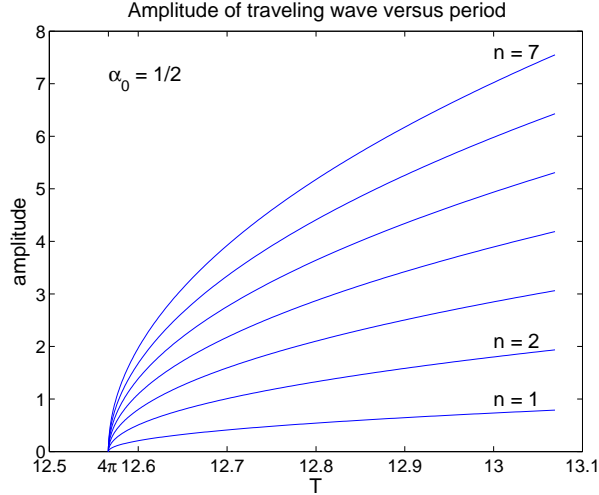}
\end{center}
\caption{First seven bifurcations from the constant solution
  $u(x)=\alpha_0$ to traveling waves with $n$-humps.  The
  period shown is $T=2\pi m/[n(n\alpha-\alpha_0)]$ with $m=2n(n-1/2)$.
  We used this to solve for $|\beta|$ in terms of $T$ via
  (\ref{eqn:alphabeta}).  The amplitude shown is the difference
  between the maximum and minimum values of the solution,
  i.e.~$8n|\beta|/(1-|\beta|^2)$.}
\label{fig:bifur:trav}
\end{figure}

We observe an analogous phenomenon when bifurcating from constant
solutions to traveling waves; see Figure~\ref{fig:bifur:trav}.  When
$u=\alpha_0$ is a constant function in (\ref{eqn:AB:def}) above,
$\hat{A}$ and $\hat{B}$ are both diagonal matrices, the eigenvalues
and eigenvectors of $BA$ are given by
$$\omega_n=n(|n|-\alpha_0), \qquad z_n(x) = e^{inx},
  \qquad (n\in\mathbb{Z}),$$
and the bifurcation times are given by
$$T_{n,m}=2\pi m/(n|\alpha_0-n|), \qquad (n,m\ge1).$$
Note that in this simplified problem, the bifurcation index $n$ turns
out to be the number of humps.  If $\alpha_0=1/2$ and $T=4\pi$, then
$\omega_n T\in2\pi\mathbb{Z}$ for \emph{every} $n$, i.e.~the kernel
$\mc{N}_1$ of
$D_1F(u,T)$ is the whole space $H^1$.  Nevertheless, the traveling
solutions that emerge from this bifurcation are no different than if
$\alpha_0$ were irrational and $\mc{N}_1$ were spanned by $\{ 1,\cos
  nx,\sin nx\}$ for some fixed $n$ --- they all just happen to join
together at $T=4\pi$ for this value of the mean.  More specifically,
the $n$-hump traveling solutions $u_{\alpha_0,n,\beta}(x,t)$ defined
in (\ref{eqn:trav:soln:def}) above have the property that as
$\beta\rightarrow0$ (and hence $\alpha\rightarrow1$), a multiple $m$
of their shortest period $2\pi/[n(n\alpha-\alpha_0)]$ converges to
$4\pi$. So while $\cos x$ and $\cos 2x$ are each bifurcation
  directions for initial conditions that lead to periodic solutions of
  the nonlinear problem, $(\cos x+\cos 2x)$ is not.

\section{The Method}
\label{method}

In order to compute non-trivial time periodic solutions, we define
the functional
\begin{equation} \label{eqn:Gtot:def}
  G_\text{tot}(u_0,T) = G(u_0,T) + \varphi(u_0,T)
\end{equation}
with
\begin{equation} \label{eqn:G:def}
  G(u_0,T) =
  \frac{1}{2}\int_0^{2\pi} [u(x,T) - u_0(x)]^2\,dx
\end{equation}
and look for minimizers of $G_\text{tot}$ with the hope that the
minimum value will be zero.  Here $\varphi(u_0,T)$ is a non-negative
penalty function designed to eliminate spatial and temporal
phase shifts and specify the mean $\alpha_0$ and the amplitude of
one of the Fourier modes at $t=0$.
Our first goal is to find an efficient method
of computing the variational derivative of $G$.  As usual in optimal
control problems \cite{pironeau:84}, there is an adjoint PDE that allows
us to compute $\varder{G}{u_0}$ in as little time as it takes to
compute $G$ itself.  We will then use a spectral method in space and a
fourth order semi-implicit Runge-Kutta method
\cite{cooper,carpenter,wilk228A} in time to solve the Benjamin-Ono and
adjoint equations to compute $G$, $\varder{G}{u_0}$ and $\der{G}{T}$
in the inner loop of the BFGS minimization algorithm \cite{bfgs,nocedal}.

\subsection{Variational Derivative of $G$}

Let $u_0$ be any function in $H^1$ (not necessarily leading to a
periodic solution).  Evidently,
\begin{equation}
  \der{}{T} G(u_0,T) = \int_0^{2\pi} [u(x,T)-u_0(x)]u_t(x,T)\,dx.
\end{equation}
Let $v_0\in H^1$ be given and define $\dot{G} = D_1G(u_0,T)v_0$, i.e.
\begin{equation} \label{eqn:G:dot}
  \dot{G} =
  \dder{}{\veps}\Big\vert_{\veps=0} G(u_0+\veps v_0,T) =
  \int_0^{2\pi} [u(x,T) - u_0(x)][v(x,T) - v_0(x)]\,dx.
\end{equation}
Here $v(x,t) = \dot{u}(x,t) = \dder{}{\veps}\big\vert_{\veps=0}
u(x,t,\veps)$ with $u(x,t,\veps)$ the solution of Benjamin-Ono with
initial condition $u(x,0,\veps)=u_0(x)+\veps v_0(x)$.  We can compute
$v$ by solving the variational equation
\begin{equation} \label{eqn:BO:lin2}
  v_t = Hv_{xx} - (uv)_x, \qquad v(x,0) = v_0(x),
\end{equation}
which is linear but non-autonomous (as $u$ depends on time in general).
Our next task is to eliminate $v(x,T)$ from (\ref{eqn:G:dot})
and represent $\dot{G}$ as an inner product:
\begin{equation} \label{eqn:G:dot2}
  \dot{G} =
  \int_0^{2\pi} \varder{G}{u_0}(x)\,v_0(x)\,dx.
\end{equation}
The idea is to define a function $w(x,s)$ going backward in time
(with $s=T-t$) such that
\begin{equation} \label{eqn:w:ic}
  w(x,0) = w_0(x) = u(x,T) - u_0(x)
\end{equation}
and then determine how $w$ should evolve so that
\begin{equation} \label{eqn:wv:integrals}
  \int_0^{2\pi} w(x,0) v(x,T) \,dx =
  \int_0^{2\pi} w(x,T) v(x,0) \, dx.
\end{equation}
Let us define the solution operator $V(t_2,t_1):H^1\rightarrow H^1$
for the linearized equation (\ref{eqn:BO:lin2}) as the mapping
that evolves an initial condition specified at time $t_1$ to the
solution at time~$t_2$.  These operators satisfy a non-autonomous,
time reversible version of familiar semigroup properties:
\begin{equation} \label{eqn:semigroup}
  V(t_1,t_1)=I, \qquad V(t_3,t_1) = V(t_3,t_2)V(t_2,t_1), \qquad
  (t_1,t_2,t_3\in\mathbb{R}).
\end{equation}
Equation (\ref{eqn:wv:integrals}) may now be written
\begin{equation}
  \left\langle w_0,V(T,0)v_0\right\rangle =
  \left\langle W(T,0) w_0, v_0 \right\rangle
\end{equation}
where $\langle\cdot,\cdot\rangle$ is the $L^2$ inner product and we
define $W(s_2,s_1) = V(t_1, t_2)^*$ with $t_j=T-s_j$.
It follows from (\ref{eqn:semigroup}) that $W(s_1,s_1)=I$ and
$W(s_3,s_1)=W(s_3,s_2)W(s_2,s_1)$.  What remains is to determine how
this non-autonomous semigroup $W$ is generated.  Taking the inner
product of $v_t$ with $w$, we have
\begin{multline}\label{wSCalc}
\int v_{t}(x,t)w(x,s)\,dx = \lim_{h\rightarrow 0}\int\left(\left[
\frac{V(t+h,t)-V(t,t)}{h}\right] v(x,t)\right)w(x,s-h)\,dx \\
=\lim_{h\rightarrow 0}\int v(x,t)\left(\left[
\frac{W(s,s-h)-I}{h}\right]w(x,s-h)\right)\,dx =
\int v(x,t)w_{s}(x,s)\,dx.
\end{multline}
We learn that
\begin{equation}\label{eqn:wIVP:derive}
  \int v w_s\, dx = \int v_t w\,dx = \int [Hv_{xx}-(uv)_x]w\,dx
  = \int v[-Hw_{xx}+uw_x]\,dx,
\end{equation}
i.e.~$w$ should solve the adjoint equation to (\ref{eqn:BO:lin2}), namely
\begin{equation}\label{eqn:wIVP}
  w_s(x,s) = -Hw_{xx}(x,s) + u(x,T-s)w_x(x,s).
\end{equation}
The time reversal in the inhomogeneous term $u(x,T-s)$ is significant.
Combining this with (\ref{eqn:G:dot}) and (\ref{eqn:G:dot2}), we
conclude that
\begin{equation}
  \frac{\delta G}{\delta u_{0}}(x) = w(x,T) - w_{0}(x),
\end{equation}
where $w$ solves (\ref{eqn:wIVP}) with initial condition (\ref{eqn:w:ic}).

\noindent{\bf{Remark:}}
We emphasize that the steps we have just followed for the
Benjamin-Ono equation can in principle be carried out for any PDE.
These steps are simply:
\begin{enumerate}
\item Find the variational equation analogous to (\ref{eqn:BO:lin})
\item Find the appropriate adjoint equation, accounting for time-reversal.
\end{enumerate}
The details of the initial condition of the adjoint problem and the
formula for $\varder{G}{u_0}$ depend on the particular functional $G$
we choose, but they are usually straightforward to work out.
For example, as another variant, we could define
\begin{equation} \label{eqn:Gsym:def}
  G(u_0,T) = \frac{1}{2}\int_0^{2\pi} [u(x,T/2)-u(2\pi-x,T/2)]^2\,dx
\end{equation}
to impose even symmetry at the half-way point.  (Recall that if
$u_0$ is symmetric, then $u(2\pi-x,T/2)=u(x,-T/2)$).  In this case
we find that
\begin{equation}
  \varder{G}{u_0}(x) = 2w(x,T/2), \qquad w_0(x) = u(x,T/2)-u(2\pi-x,T/2),
\end{equation}
or, since $v_0$ is assumed symmetric in this formulation,
$\varder{G}{u_0}(x)=w(x,T/2)+w(2\pi-x,T/2)$.  In subsequent work, we
will apply the methods of this paper to the vortex sheet with surface
tension and to the water wave.  Although step 2 usually amounts
  to a simple integration by parts as was done in
  (\ref{eqn:wIVP:derive}) above, the adjoint calculation can be quite
  involved for systems of PDEs with more complicated nonlinearities
  such as the Birkhoff-Rott integral in the vortex sheet problem
  (see \cite{ambrose:wilkening:vtx}).

\subsection{The Numerical Method} \label{sec:numerical:method}

We minimize $G_\text{tot}$ using the BFGS algorithm \cite{nocedal},
which is a quasi-Newton line search method that builds an approximate
Hessian incrementally from the history of gradients it has evaluated.
As a black box unconstrained minimization algorithm, it requires only
an initial guess and subroutines to compute $G_\text{tot}(q)$ and
$\nabla_qG_\text{tot}(q)$, where $q\in\mathbb{R}^{d}$ contains the
numerical degrees of freedom used to represent $u_0$ and $T$.  We use
a solution of the linearized problem for the initial guess near a
stationary solution or traveling wave, and then use linear
extrapolation (or the result of the previous iteration) for the
initial guess in subsequent calculations as we vary the bifurcation
parameter.

In our implementation, we wrote a C++ wrapper around J. Nocedal's
L-BFGS Fortran code released in 1990, but we turn off the limited
memory aspect of the code since computing $G$ takes more time than the
linear algebra associated with updating the full Hessian matrix.  We
do find that the algorithm converges quadratically once it gets close
to a minimizer.  Our code also makes use of the FFTW and LAPACK
libraries, but was otherwise written from scratch.

We represent $u(x,t)$ spectrally as a sum of $M$ (typically 384 or 512)
Fourier modes,
\begin{equation}
  u(x,t) = \sum_{k=-M/2+1}^{M/2} c_k(t) e^{ikx}, \qquad c_k\in\mathbb{C}.
\end{equation}
Since $u$ is real, we use the r2c version of the FFT algorithm, which
only accesses the coefficients $c_k$ with $k\ge0$, assuming
$c_{-k}=\bar{c}_k$.  We also zero out the Nyquist frequency $c_{M/2}$
so that the total number of (real) degrees of freedom representing $u$
at time $t$ is $M-1$.  We use $d=M/2$ degrees of freedom to represent
$u_0$ and $T$, namely
\begin{equation} \label{eqn:q:def}
  q=(a_0,T,a_1,b_1,\dots,a_{M/4-1},b_{M/4-1})\in\mathbb{R}^d,
  \qquad (c_k = a_k + ib_k, \; t=0).
\end{equation}
The remaining Fourier modes in $u_0$ are taken to be zero.  The reason
for using fewer Fourier modes in the initial condition is that in
order to avoid aliasing errors, we want the upper half of the spectrum
to remain close to zero throughout the calculation; therefore, we do
not wish to give BFGS the opportunity to modify these coefficients.
We increase $M$ and repeat the calculation any time one of the high
frequency ($k\ge M/4$) Fourier modes of the optimal solution exceeds
$10^{-13}$ in magnitude at any timestep.

To compute $G(q)$, we write the Benjamin-Ono equation in the form
\begin{equation}
  u_t = f(u) + g(u), \qquad g(u) = Hu_{xx}, \quad
  f(u) = - \left(\jt\frac{1}{2}u^2\right)_x,
\end{equation}
where $\frac{1}{2}u^2$ is evaluated on the grid $\{x_j=2\pi
j/M\;:\;0\le j\le M-1\}$ in physical space while $H$ and $\partial_x$
are evaluated in Fourier space.  The trapezoidal rule in physical
space is used to evaluate the integral (\ref{eqn:G:def}) defining $G$.
To evolve the solution, we use the stiffly stable, additive
(i.e.~implicit-explicit) Runge-Kutta method of Kennedy and Carpenter
\cite{carpenter,wilk228A} known as ARK4(3)6L[2]SA with a fixed
timestep $h=T/N$, where $N$ is chosen to be large enough that further
refinement does not improve the solution.  Briefly, the idea of an ARK
method is to treat $f$ explicitly (as it is non-linear) while treating
$g$ implicitly (as it is the source of stiffness):
\begin{equation}
  \begin{aligned}[c]
    &k_i = f\left(\jt t_n + c_ih, \, u_n + h\sum_j a_{ij}k_j +
      h \sum_j \hat{a}_{ij}\ell_j\right), \\
    &\ell_i = g\left(\jt t_n + \hat{c}_ih, \, u_n + h\sum_j a_{ij}k_j +
      h\sum_j \hat{a}_{ij}\ell_j\right), \\
    &u_{n+1} = \jt u_n + h\sum_j b_j k_j + h\sum_j \hat{b}_j \ell_j.
  \end{aligned} \qquad
  \parbox[c][1.1in][t]{.6in}{
  $$\begin{array}{c|c}
    c & A \\ \hline & b^T
  \end{array}$$
  \centering for $f$} \qquad
  \parbox[c][1.1in][t]{.6in}{
  $$\begin{array}{c|c}
    \hat{c} & \hat{A} \\ \hline & \hat{b}^T
  \end{array}$$
  \centering for $g$}
\end{equation}
The Butcher array for $f$ satisfies $a_{ij}=0$ if $i\le j$ and for $g$
satisfies $\hat{a}_{ij}=0$ if $i<j$, which allows the stage
derivatives to be solved for in order: $\ell_1$, $k_1$, $\ell_2$,
$k_2$, \dots, $\ell_6$, $k_6$, where our scheme has $6$ stages.  See
\cite{carpenter} for the scheme coefficients and \cite{wilk228A} for
details on solving the implicit equations in the similar cases of
Burgers' equation and the KdV equation.

Once $u(x,T)$ is known, we use the same scheme to solve the adjoint
equation
\begin{equation} \label{eqn:split:adjoint:eqn}
  w_s = f(s,w) + g(w), \qquad g(w) = -Hw_{xx}, \qquad
  f(s,w)(x) = u(x,T-s)w_x(x).
\end{equation}
The main difficulty is that the intermediate stages of the ARK method
require the value of $u$ at intermediate times (between timesteps).
For this we use cubic Hermite interpolation, matching $u$ and $u_t$
at the timesteps straddling the required intermediate time:
\begin{equation*}
  u(\cdot,t_n+\theta h) = (1-\theta)u_n + \theta u_{n+1} -
  \theta(1-\theta)\big[ (1-2\theta)(u_{n+1}-u_n)-
  (1-\theta)h\partial_tu_n + \theta h \partial_t u_{n+1}\big]
\end{equation*}
where $0<\theta<1$.  This yields fourth order accurate values of
$u$ in the right hand side of (\ref{eqn:split:adjoint:eqn}),
which is sufficient to achieve a fourth order accurate global
solution $w$.  We include the option in our code to store $u$
only at certain milemarker times, and then regenerate the
data at all timesteps between milemarkers as soon as the $w$
equation enters that region; this dramatically reduces the memory
requirements of the code at the expense of having to compute
$u$ twice.

Once $u(x,T)$ and $w(x,T)$ are known with the period and initial
conditions specified in $q\in\mathbb{R}^d$, we compute
$G(q)$ using the trapezoidal rule in physical space to evaluate the
integral in (\ref{eqn:G:def}), and we compute $\der{G}{q_j}$
by taking the FFT of $\varder{G}{u_0}$ and scaling each component
appropriately:
\begin{equation} \label{eqn:dGdq}
\begin{aligned}
  \der{G}{q_0} &= \int_0^{2\pi} \varder{G}{u_0}(x)\,1\,dx =
  2\pi\left(\jt \varder{G}{u_0}\right)^\wedge_0 \\[2pt]
  \der{G}{q_1} &= \der{G}{T} =  \int_0^{2\pi} [u(x,T)-u_0(x)]u_t(x,T)\,dx,
  \quad\; \longleftarrow \;\quad \left(\parbox{1.2in}{\centering
    use trap.~rule\\ in physical space}\right) \\[2pt]
  \der{G}{q_{2k}} &= \der{G}{a_k} =
  \int_0^{2\pi} \varder{G}{u_0}(x)\left( e^{ikx} + e^{-ikx} \right)dx =
  4\pi\real\left\{\left(\jt \varder{G}{u_0} \right)^\wedge_k\right\},
  \hspace*{.34in} (k\ge1), \\[2pt]
  \der{G}{q_{2k+1}} &= \der{G}{b_k} =
  \int_0^{2\pi} \varder{G}{u_0}(x)\left( ie^{ikx} - ie^{-ikx} \right)dx =
  4\pi\imag\left\{\left(\jt \varder{G}{u_0} \right)^\wedge_k\right\},
  \hspace*{.25in} (k\ge1).
\end{aligned}
\end{equation}
We remark that these formulas for the derivatives of the numerical
version of $G$ essentially assume that we have solved the PDE exactly
(so that the calculus of variations applies to our numerical
solutions).  This is reasonable in our case as we are using spectrally
accurate schemes, but would cause difficulties if the numerical
solution were only first or second order accurate in space or time.

\subsection{Choice of Penalty Function $\varphi$}

We still need to define the penalty function $\varphi(u_0,T)$ in
(\ref{eqn:Gtot:def}) and show how to compute its gradient with respect
to $q$.  The purpose of $\varphi$ is to pin down the mean and the
phase shifts in space and time as well as to specify the bifurcation
parameter.  We have explored several successful variants which became
more specialized as our understanding of the problem increased.  As
some of these variants may prove useful in other problems, we describe
them here.

Initially we did not include a penalty function in $G_\text{tot}$, but
without it, the BFGS algorithm invariably converges to a constant
solution.  Next we constrained $q_2$, the real part of the first
Fourier mode $\hat{u}_1(t)=a_1(t)+ib_1(t)$ at $t=0$, to have a given
value $\rho$.  We reasoned that as long as $\rho$ is not too
large, the BFGS algorithm can vary $q_3=b_1(0)$ to find a periodic
solution, so all we are doing is pinning down a phase.  This was done
by defining
\begin{equation*}
  \varphi(u_0,T) = \frac{1}{2}\Big(
  [a_0(0)-\alpha_0]^2 +
  [a_1(0)-\rho]^2\Big)
  \qquad
  \text{or} \qquad
  \varphi(q) = \frac{1}{2}\Big(
  [q_0-\alpha_0]^2 +
  [q_2-\rho]^2 \Big),
\end{equation*}
which works well to rule out the constant solutions but generally
leads to traveling waves.  By studying these traveling waves, we
determined the formulas of Section~\ref{stationary} and also observed
that for some choices of $\rho$ and starting guess $q^\e0$, the wave
becomes ``wobbly,'' indicating that a non-trivial solution might be
nearby.

To rule out traveling waves, we chose a parameter $\eta\in[-1,1]$ and
defined
\begin{equation*}
  \varphi(u_0,T) = \frac{1}{2}\Big(
  [a_0(0)-\alpha_0]^2 +
  [a_1(0) - \rho]^2 +
  [a_1(T/2) - \eta a_1(0)]^2 +
  [b_1(T/2) - \eta b_1(0)]^2\Big).
\end{equation*}
Our idea here was that a (one-hump) traveling
wave would have $\eta=\pm1$, depending on how many times it passed
through the domain in time $T$; hence, intermediate values of $\eta$
would have to correspond to non-trivial solutions.  To compute
the gradient of $\varphi$ when it involves Fourier modes at later
times, we simply solve another adjoint problem.  Specifically, if
$\varphi$ involves one of
\begin{equation*}
  a_k(T/2) = \frac{1}{2\pi}\int_0^{2\pi} u(x,T/2)\cos(kx)\,dx, \qquad
  b_k(T/2) = \frac{1}{2\pi}\int_0^{2\pi} u(x,T/2)[-\sin(kx)]\,dx,
\end{equation*}
we will need to compute $\varder{}{u_0}a_k(T/2)$ or
$\varder{}{u_0}b_k(T/2)$, which can be done by setting
\begin{equation*}
  w_0(x) = \frac{1}{2\pi}\cos(kx), \qquad \text{or} \qquad
  w_0(x)=-\frac{1}{2\pi}\sin(kx)
\end{equation*}
and solving (\ref{eqn:wIVP}) from $s=0$ to $s=T/2$; the result
$w(x,T/2)$ is the desired variational derivative.  These may then be
used to compute $\der{}{q_j}a_k(T/2)$ or $\der{}{q_j}b_k(T/2)$ as was
done for $G$ in (\ref{eqn:dGdq}), at which point it is a simple matter
to obtain $\der{\varphi}{q_j}$.

This procedure proved very effective in obtaining non-trivial time
periodic solutions.  The BFGS algorithm is able to minimize
$G_\text{tot}$ down to $10^{-26}$, at which point roundoff error
prevents further reduction.  With random initial data $q^\e0$ (which
we tried before we had solved the eigenvalue problem for the
linearization), the algorithm explores quite a wide region of the
parameter space, with all components of $q$ (including $T$) changing
substantially --- we do not seem to get stuck in non-zero local minima
of $G_\text{tot}$.  Once we do find a nontrivial solution, varying
$\eta$ leads to other nearby periodic solutions.

Studying this family of solutions, we finally realized that we were
dealing with a four parameter family of nontrivial solutions with the
mean, two phases and a bifurcation parameter describing them.  The
main drawback of using $\eta$ as the bifurcation parameter is that the
spatial and temporal phases are not specified independently, but
instead depend on $\eta$ in a complicated way.  A more natural choice
is to define
\begin{equation} \label{eqn:phi}
  \varphi(u_0,T) = \frac{1}{2}\Big(
  [a_0(0) - \alpha_0]^2 + [a_k(0) - \rho]^2 + [b_k(0)]^2 +
  [\partial_t a_k(0)]^2\Big),
\end{equation}
i.e.~we use $\varphi$ to impose the mean $\alpha_0$, the bifurcation
parameter $\rho$, the spatial phase $b_k(0)=0$, and the temporal
phase $\partial_ta_k(0)=0$.  Given any solution, we can always
translate space and time to achieve the latter two conditions --- we
have not made any symmetry assumptions here.  The index $k$ we use
depends on the number of humps $N$ and bifurcation index $n$ of
the linearized solution; the only requirement is that $\hat{z}_{N,n,k}$
in (\ref{eqn:evec:formulas}) be non-zero.  One readily checks that
\begin{equation}
  \partial_ta_k(0) = \frac{1}{2\pi}\int_0^{2\pi} u_t(x,0)\cos kx\,dx =
  \frac{1}{2\pi}\int_0^{2\pi}
  \left[-k^2 u_0 + (k/2)u_0^2\right](-\sin kx)\,dx,
\end{equation}
from which we obtain $\varder{}{u_0}[\partial_ta_k(0)](x) =
\frac{1}{2\pi}(k^2 - k u_0(x))\sin kx$.  Although (\ref{eqn:phi}) does
not rule out traveling waves, we have no difficulty bifurcating from
traveling waves to non-trivial solutions by choosing a starting
guess that includes first order corrections from the linear theory
of Section~\ref{linearTheory}.

\section{Non-Trivial Time-Periodic Solutions}\label{sec:nontrivial}

We now use the methods described above to study the global behavior of
non-trivial time-periodic solutions far beyond the realm of validity
of the linearization about stationary and traveling waves.  We find
that these non-trivial solutions act as rungs in a ladder, connecting
stationary and traveling solutions with different speeds and
wavelengths by creating or annihilating oscillatory humps that grow or
shrink in amplitude until they become part of the stationary or
traveling wave on the other side of the rung.  The dynamics of these
non-trivial solutions are often very interesting, sometimes resembling
a low amplitude traveling wave superimposed on a lower frequency
carrier signal, and other times behaving like two bouncing solitons
that repel each other to avoid coalescing.  In this section, we
present a detailed numerical study of the path of non-trivial
solutions connecting the one-hump stationary solution to the two-hump
traveling wave.  In Section~\ref{sec:exact}, we derive exact formulas
for the solutions on this path.  In a follow-up paper
\cite{benj:ono2}, we classify all bifurcations from traveling waves,
study the paths of non-trivial solutions connecting several of them,
and propose a conjecture explaining how they all fit together.

\begin{figure}
\begin{center}
\includegraphics[width=.9\linewidth]{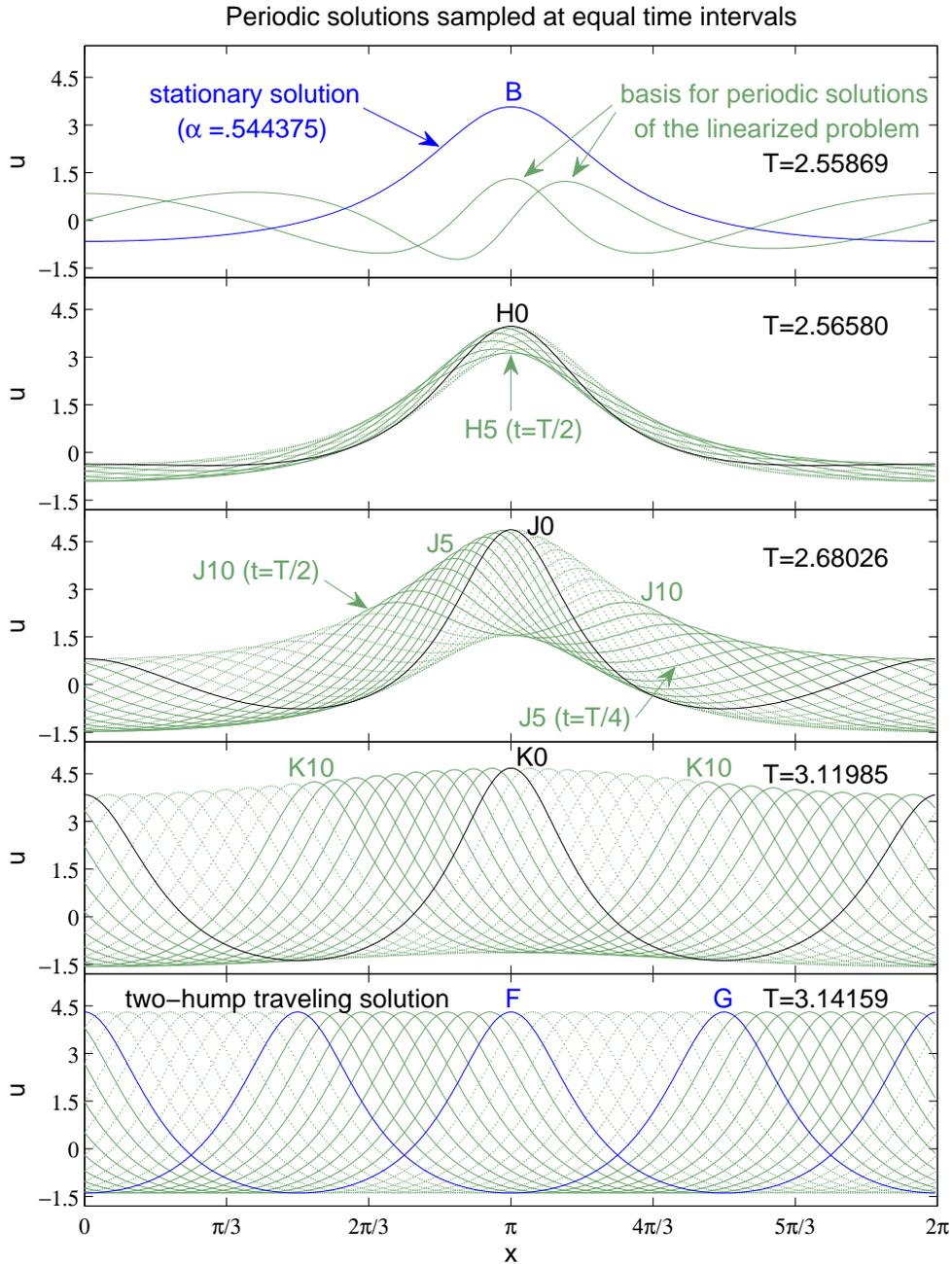}
\end{center}
\caption{Progression from a one-hump stationary solution (top) to a
  two-hump traveling wave (bottom, moving left) by varying the real
  part of the first Fourier mode at $t=0$ while holding the mean
  $\alpha_0$ constant and choosing the spatial and temporal phases
  such $\varphi$ in (\ref{eqn:phi}) is zero.  The labels B, F, G, H0,
  J0, etc.~correspond to
  Figures~\ref{fig:bifur1}--\ref{fig:beta:two:hump}.}
\label{fig:evol}
\end{figure}

\begin{figure}[p]
\begin{center}
\includegraphics[width=.692\linewidth,trim=0 20 0 0]{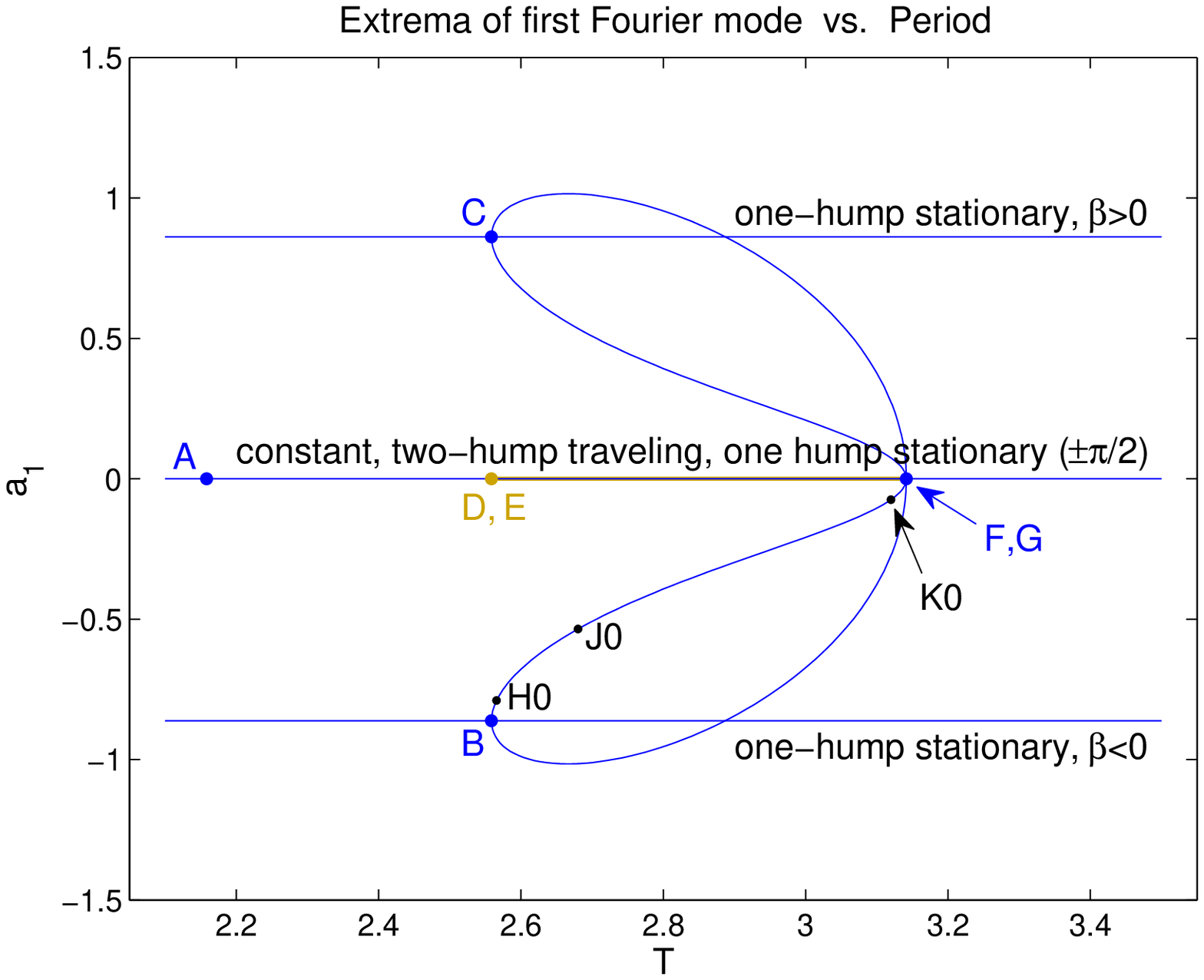}
\end{center}
\caption{ Bifurcation from one-hump stationary solutions (B and C) to
  non-trivial time-periodic solutions that re-connect with two-hump
  traveling waves at F and G.}
\label{fig:bifur1}
\end{figure}

Consider the periodic solutions obtained by bifurcation from the
one-hump stationary solution at the lowest frequency, $\omega_{1,1}$.
We arbitrarily set the mean $\alpha_0=0.544375$ for these simulations
(see Figure~\ref{fig:stationary} above), but as shown in
Section~\ref{sec:exact}, any choice of $\alpha_0<1$ would lead
to similar results.  In the top pane of Figure~\ref{fig:evol}, we show
the one-hump stationary solution $u_{1,\beta}(x)$ with
$\beta=-\sqrt{(1-\alpha_0)/(3-\alpha_0)}$ together with the (initial
conditions of the) two periodic solutions
\begin{equation}
  v^\e{0}(x,t) = \real\{z_{1,1}(x)e^{i\omega_1 t}\}, \qquad
  v^\e{1}(x,t) = \imag\{z_{1,1}(x)e^{i\omega_1 t}\}
\end{equation}
of the linearized equation (\ref{eqn:BO:lin}) corresponding to
the first eigenvalue $\omega_{1,1} = 3 - \alpha_0$ of
$BA=-i\partial_x(H\partial_x-u)$.  The natural period of these
solutions is $T=2\pi/\omega_{1,1} = 2.55869$.  Note how the non-linearity
of Benjamin-Ono distorts these two-hump perturbations as they travel
(to the left) on top of the one-hump stationary ``carrier'' solution.
Also note that $v^\e0$ and $v^\e1$ are actually the same solution
with a $T/4$ phase lag in time:
\begin{equation*}
  v^\e0(x,T/4)=-v^\e1(x,0) \quad \text{while} \quad
  v^\e1(x,T/4)= v^\e0(x,0).
\end{equation*}
We choose the real part of the first Fourier mode as the bifurcation
parameter $\rho$ so that $k=1$ in the definition (\ref{eqn:phi}) of
$\varphi$.  As we vary $\rho=a_1(0)$ from
$-2\sqrt{(1-\alpha_0)/(3-\alpha_0)}$ to $0$, we traverse the
trajectory from B to F in the bifurcation diagram of
Figure~\ref{fig:bifur1}.  The curves corresponding to the intermediate
points H0, J0 and K0 along this path are shown in black in panes 2--4
of Figure~\ref{fig:evol}.  Along this path, we see that a second hump
forms at $x=0$ while the center hump sharpens to accommodate the
shorter wavelength of the two-hump traveling wave.  If we instead
increase $|\rho|$ near point B in the diagram, we obtain the lower
path connecting B to G.  Along this path, the center hump decreases in
magnitude (curve H5), forms a dimple in the middle (curve J10), splits
into two humps (curve K10), and again turns into a two-hump traveling
wave (curve G).  These curves are related to those on the path from B
to F by a $T/2$ phase shift in time.  If we change the sign of $\beta$
(i.e.~shift the phase by $\pi$) in the stationary solution and call
the resulting curve C, the bifurcation diagram is reflected about the
$T$ axis.  The path from B (or C) to F is easier to compute due to the
turning point in $|\rho|$ on the path from B (or C) to G.

\begin{figure}[p]
\begin{center}
\includegraphics[width=.87\linewidth,trim=5 20 0 0]{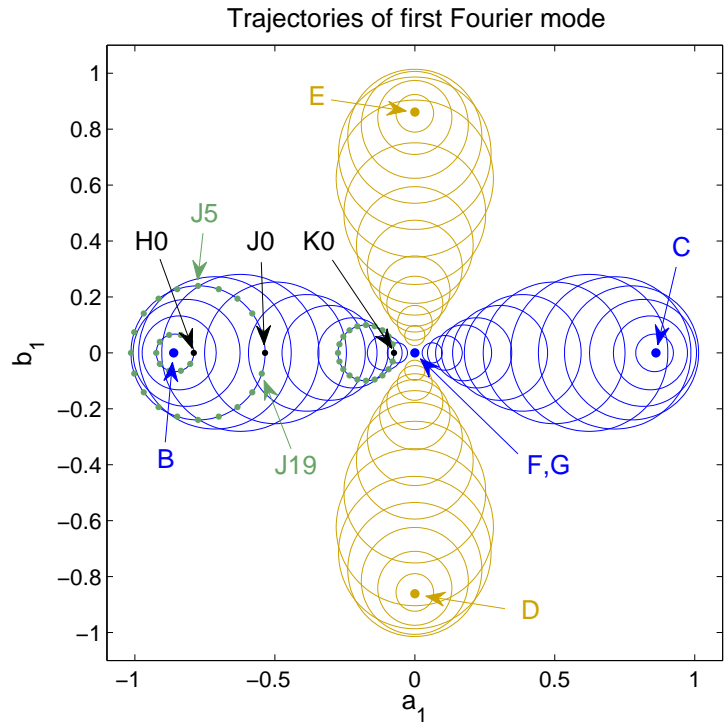}
\end{center}
\caption{
The trajectories of the first Fourier mode in the complex plane are
exactly circular.  The markers on the left lobe correspond to the
solutions shown in Figure~\ref{fig:evol}.
}
\label{fig:circles1}
\end{figure}

By the time we reach $K0$ on the path from B to F, we can view our
solution as a two-hump traveling wave with a small one-hump stationary
perturbation corresponding to the first eigenvalue $\omega_{2,1}=1$ in
the linearization about the two-hump traveling wave.  A full analysis
of the linearization about traveling waves is given in the follow-up
paper \cite{benj:ono2}, but the idea is that if $u(x)$ is a stationary
solution and $U(x,t)=u(x-ct)+c$ is a traveling wave, then the
solutions $v$ and $V$ of the linearizations about $u$ and $U$ satisfy
$V(x,t)=v(x-ct,t)$.  Now, the linearized solutions
$\real\{z_{2,1}(x)e^{i\omega_{2,1}t}\}$ and
$\imag\{z_{2,1}(x)e^{i\omega_{2,1}t}\}$ about the two-hump
\emph{stationary} solution have the property that
$z_{2,1}(x-\pi)=-z_{2,1}(x)$; hence, when they are used as
perturbations on a two-hump \emph{traveling} wave, they need to
progress through an extra half-cycle in time to make up for the sign
change.  As a result, $\omega_{2,1}T$ must belong to
$\pi+2\pi\mathbb{Z}$ (rather than $2\pi\mathbb{Z}$ itself) for the
linearized solution to be periodic.  It turns out that as we traverse
the path from B to F, the period of the solution increases from
$T=2\pi/\omega_{1,1}$ up to $T=\pi/\omega_{2,1}=\pi$ (rather than
e.g.~$3\pi$ or $5\pi$).  Note that as $\omega_{2,1}=1$ is
independent of $\alpha_0$, the path connecting the one hump stationary
solution to the two-hump traveling wave always terminates at $T=\pi$,
regardless of the mean.

In Figure~\ref{fig:circles1}, we plot the trajectories of the first
Fourier mode $c_1(t)=a_1(t)+ib_1(t)$ in the complex plane for various
choices of the bifurcation parameter $\rho=a_1(0)$.  We were
surprised to find that these trajectories are exactly circular; this
will be discussed further below.  The markers on the left (west) lobe
of circles correspond to solutions plotted in Figure~\ref{fig:evol};
for example, J19 corresponds to $u\left(x,\frac{19}{20}T\right)$,
which is the dotted curve immediately to the right of the initial
condition J0 in the center pane of Figure~\ref{fig:evol}.  For
visibility, we only plotted 10 timeslices in the evolution of H0.

The four parameter family of non-trivial solutions can be seen
in Figure~\ref{fig:circles1}.  A given solution is
represented by one of the circular trajectories.  The two main
parameters describing this family are the mean $\alpha_0$ and the
distance from the nearest point on the circle to the origin.  A
spatial phase shift of the initial condition by $\theta$ (with the
sign convention of Eq.~(\ref{eqn:S:def})) amounts to a clockwise
rotation of the circle about the origin by $\theta$ (or $k\theta$ for
the $k$th Fourier mode).  The north, east and south lobes of circles
represent spatial phase shifts of the
west lobe of solutions by $\theta=\pi/2$, $\pi$ and $-\pi/2$,
respectively, but any other phase shift $\theta\in\mathbb{R}$ is also
allowed.  Finally, a temporal phase shift amounts to choosing which
point on the circle we assign to $t=0$.  Requiring that the initial
condition have even symmetry yields either the west or east lobe of
solutions with $t=0$ occurring along the real axis.

\begin{figure}[p]
\begin{center}
\includegraphics[width=.65\linewidth,trim=0 20 0 0]{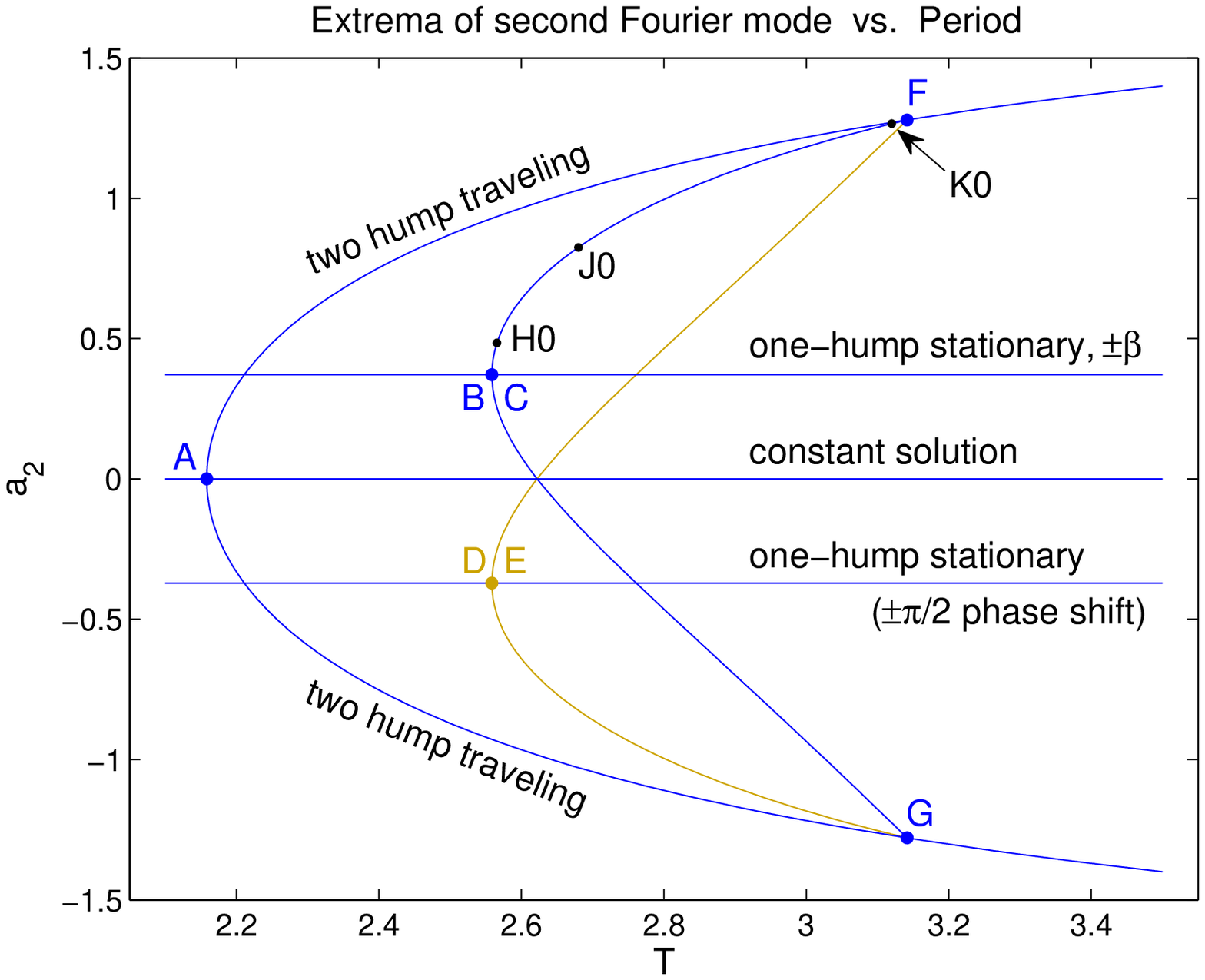}
\end{center}
\caption{Bifurcation from the constant solution to a two-hump traveling
wave and the path of non-trivial solutions connecting these to various
one-hump stationary solutions.}
\label{fig:bifur2}
\end{figure}

We can also use other Fourier modes for the bifurcation parameter.
This is especially important to track higher order bifurcations from
multi-hump traveling waves --- in these cases, the first several
Fourier modes remain zero for all solutions in the family at all
times~$t$.  But even for the simplest path connecting one-hump
stationary solutions to two-hump traveling waves, it is useful to
study other bifurcation diagrams representing this same family of
solutions.  In Figure~\ref{fig:bifur2}, we show the result when the
second Fourier mode is used instead of the first.  By setting
$\rho=a_2(0)$, we can now also see the bifurcation (labeled A) from
the constant solution $u\equiv\alpha_0$ to the two-hump traveling
waves; moreover, the points F and G that fell on top of each other in
Figure~\ref{fig:bifur1} become distinct.  The outer curve connecting F
to G via A represents the set of two-hump traveling waves moving left
with mean $\alpha_0$.  This curve was plotted parametrically, setting
$a_2=\pm 2N\sqrt{(1-\alpha)/(3-\alpha)}$ and
$T=2\pi/[N(N\alpha-\alpha_0)]$ with $N=2$ and $\alpha$ ranging over
all values such that $\alpha\le1$ and $T\le3.5$.

\begin{figure}[p]
\begin{center}
\includegraphics[width=.82\linewidth,trim=3 20 0 0]{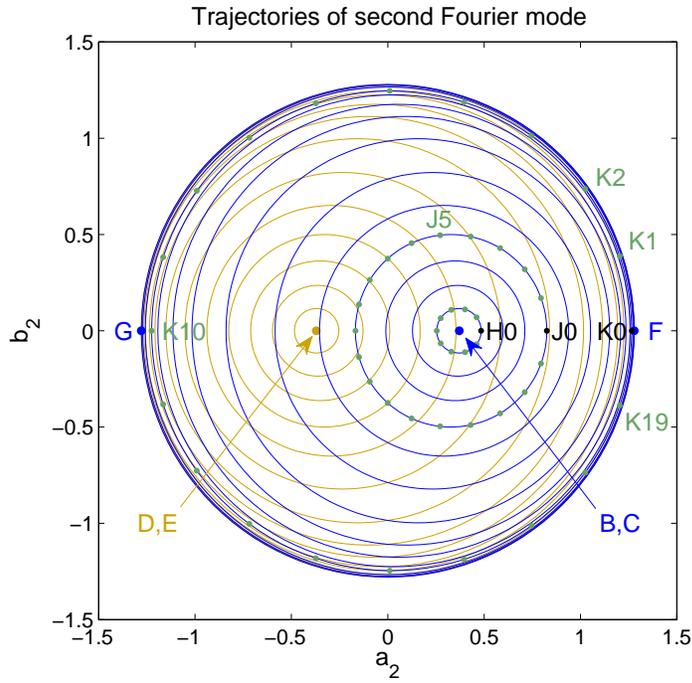}
\end{center}
\caption{The trajectories of the second Fourier mode in the complex
plane are epitrochoids; see Equation (\ref{eqn:c2:epitrochoid}).
The markers correspond to the solutions plotted in Figure~\ref{fig:evol}.
}
\label{fig:circles2}
\end{figure}

It is interesting to note that the bifurcation at F (and at G) from
the two-hump traveling wave does
not look like a pitchfork.  Instead, the bifurcation
curve enters at an oblique angle from one side only.  This is because
the second Fourier mode of the linearized solution
$v^\e0(x,t)=\real\{z_{2,1}(x+t)e^{it}\}$ is zero
(cf.~(\ref{eqn:evec:formulas})~above), so the first order effect on
the bifurcation parameter $\rho=a_2(0)$ is zero as we move away from
the two-hump traveling wave in the direction of $v^\e0$.  The
derivative of $T$ in this direction is also zero, so a
heuristic argument based on the Liapunov-Schmidt reduction
\cite{golubitsky,kielhofer} leads to an equation $g(\rho,T)=0$
for the bifurcation curve, where $\der{g}{T}=0$ and $\der{g}{\rho}=0$
at the bifurcation.
By contrast, the first Fourier coefficient of $v^\e0(\cdot,0)$ is
non-zero and we do obtain a pitchfork bifurcation at F (and G) when we
plot $a_1(0)$ vs.~$T$, as was seen in Figure~\ref{fig:bifur1}.

It turns out that the path of $a_2(0)$ vs.~$T$ from F to B is
identical to the one from F to~C; which one-hump stationary solution
we end up with depends on whether we perturb the traveling wave in the
direction of $+v^\e0$ or $-v^\e0$.  However, there is another
direction we can move while keeping $G_\text{tot}$ zero (with $k=2$ in
(\ref{eqn:phi})), namely $v^\e1(x,t)=\imag\{z_{2,1}(x+t)e^{it}\}$.
This direction breaks the even symmetry of the initial condition, but
the \emph{even} Fourier modes still satisfy $b_{k}(0)=0$ and $\partial_t
a_{k}(0)=0$; hence, the penalty function $\varphi$ does not exclude
this direction when $k=2$ in (\ref{eqn:phi}).  Depending on whether we
perturb in the $+v^\e1$ or $-v^\e1$ direction, we end up at either the
one-hump stationary solution E, with maximum at $x=3\pi/2$, or D, with
maximum at $x=\pi/2$.  This shows that our choice of penalty function
$\varphi$ in (\ref{eqn:phi}) does not rule out non-trivial solutions
with asymmetric initial conditions: the solutions on the path from
(F or G) to (D or E) are all asymmetric at $t=0$; however, these solutions
are related to the ones on the path from (F or G) to (B or C) by a phase
shift in space.  We have not found any periodic solutions that cannot
be made symmetric at $t=0$ by such a phase shift.

In Figure~\ref{fig:circles2}, we show the trajectories of the second
Fourier mode in the complex plane.  The markers labeled H0, J0,
etc.~again correspond to the solutions plotted in
Figure~\ref{fig:evol}.  Unlike the first Fourier mode, these
trajectories are not exactly circular --- but by curve fitting
we determined they are epitrochoids, (resembling Ptolemy's model
of planetary motion, or ``spirograph'' trajectories), of the form
\begin{equation} \label{eqn:c2:epitrochoid}
  c_2(t) = c_{20} + c_{2,-1}e^{i\omega t} + c_{2,-2}e^{i2\omega t},
  \qquad \left(\omega = 2\pi/T\right),
\end{equation}
where the coefficients $c_{2j}$ (and $\omega$) depend on the
bifurcation parameter $\rho$.  More generally, by curve fitting our
numerical solutions, we have discovered a rather amazing property of
solutions on this path: the $k$th Fourier mode is found to be of the
form
\begin{equation} \label{eqn:ck:form}
  c_k(t) = \sum_{j=-k}^0 c_{kj}e^{-ij\omega t}, \qquad (k\ge0, \;
  \omega = 2\pi/T),
\end{equation}
where $c_{kj}\in\mathbb{R}$ and $c_{-k}(t)=\overline{c_k(t)}$.  The
general form of solutions on other paths connecting higher order
bifurcations is similar, and is described in the follow-up paper
\cite{benj:ono2}.  The four parameter family of non-trivial solutions
is also nicely represented in this figure, where the parameters are
the mean, the furthest point on the epitrochoid, a global rotation
about the origin, and the choice of which point on the epitrochoid
corresponds to $t=0$.  Note that a spatial phase shift of the initial
condition by $\theta$ leads to a rotation of a trajectory in this
figure clockwise by $2\theta$, so the north and south lobes of circles
in Figure~\ref{fig:circles1} collapse onto the west family of
epitrochoids (around D and E) in Figure~\ref{fig:circles2} while the west
and east lobes of Figure~\ref{fig:circles1} collapse onto the east
family here.

\section{Exact Solutions} \label{sec:exact}

The discovery that the Fourier modes execute Ptolemaic orbits of the
form (\ref{eqn:ck:form}) led us to expect that it might be possible to
write down the solution in closed form.  In this section, we show how
to do this for the path of non-trivial solutions connecting the
one-hump stationary solution to the two-hump traveling wave.  We have
now learned of several other methods for finding exact solutions of
Benjamin-Ono, notably the bilinear formalism used by Satsuma and
Ishimori \cite{satsuma:ishimori:79} and Matsuno \cite{matsuno:04} to
construct multi-periodic solutions; the reduction by Case
\cite{case:mero} of the ODE (\ref{eqn:soliton:ode}) to a system shown
by Moser to be completely integrable; and the approach of Dobrokhotov
and Krichever \cite{dobro:91} using the theory of finite zone
integration to construct multi-phase solutions.  Our approach
highlights a feature of these solutions that has not been discussed
previously, namely that the Fourier modes of these solutions turn out
to be power sums of particle trajectories $\beta_l(t)$ in the unit
disk $\Delta\subset\mathbb{C}$ whose elementary symmetric functions
execute simple circular orbits in the complex plane.

We start with the observation that the meromorphic solutions
\begin{equation} \label{eqn:beta:rep}
  u(x,t) = \alpha_0 + \sum_{l=1}^N u_{\beta_l(t)}(x), \qquad
  \beta_l(t)\in\Delta \text{ satisfying (\ref{eqn:beta:ode}),}
\end{equation}
have the property that the first $N+1$ Fourier modes $c_k(t)$ of
$u(x,t)$ are closely related to the trajectories of the $\beta_l$.
Specifically, $\alpha_0=c_{0}$ is needed to write down the ODE
(\ref{eqn:beta:ode}), and we have
\begin{equation}
  \begin{aligned}
    \beta_1(t) + &\cdots + \beta_N(t) = s_1(t), & 2s_1(t)&=c_1(t), \\
    \beta_1^2(t) + &\cdots + \beta_N^2(t) = s_2(t), & 2s_2(t)&=c_2(t), \\
    &\cdots \\
    \beta_1^N(t) + &\cdots + \beta_N^N(t) = s_N(t), &\qquad 2s_N(t)&=c_N(t).
  \end{aligned}
\end{equation}
It is a standard theorem of algebra \cite{waerden1} that the
elementary symmetric functions
\begin{equation}
  \sigma_j = \sum_{l_1<\cdots<l_j} \beta_{l_1}\cdots\beta_{l_j}, \qquad
  (j=1,\dots,N)
\end{equation}
are polynomials in the power sums, e.g.
\begin{equation}
  \sigma_0 = 1, \qquad
  \sigma_1 = s_1, \qquad
  \sigma_2=\frac{s_1^2-s_2}{2}, \qquad
  \sigma_3=\frac{s_1^3-3s_1s_2+2s_3}{6}.
\end{equation}
The general recurrence relation is
\begin{equation} \label{eqn:recurrence}
  \sigma_0 = 1, \qquad
%    (-1)^j j\sigma_j + \sum_{k=0}^{j-1}(-1)^k\sigma_k s_{j-k} = 0,
  \sigma_j = \frac{1}{j}\sum_{l=1}^j (-1)^{l-1}\sigma_{j-l}s_l,
    \qquad  (j=1,\dots,N).
\end{equation}
The $\beta_l$ are then the zeros of the polynomial
\begin{equation} \label{eqn:poly}
  \prod_{l=1}^N(z-\beta_l(t)) = \sum_{j=0}^N (-1)^j\sigma_j(t) z^{N-j}.
\end{equation}
We can test whether a given numerical solution $u(x,t)$ is an
$N$-particle meromorphic solution by computing its first $N+1$ Fourier
coefficients $c_k(0)=2s_k(0)$, using (\ref{eqn:recurrence}) to obtain
the symmetric functions $\sigma_j(0)$, solving for the roots
$\beta_l(0)$ of the polynomial on the right hand side of
(\ref{eqn:poly}), and checking that higher power sums do in fact agree
with the Fourier coefficients of the solution:
\begin{equation}
  \beta_1^k(0)+\cdots+\beta_N^k(0) = \frac{1}{2}c_k(0), \qquad
  (k\ge N+1).
\end{equation}

\begin{figure}[t]
\begin{center}
  \includegraphics[height=2.75in,trim=10 0 0 0,clip]{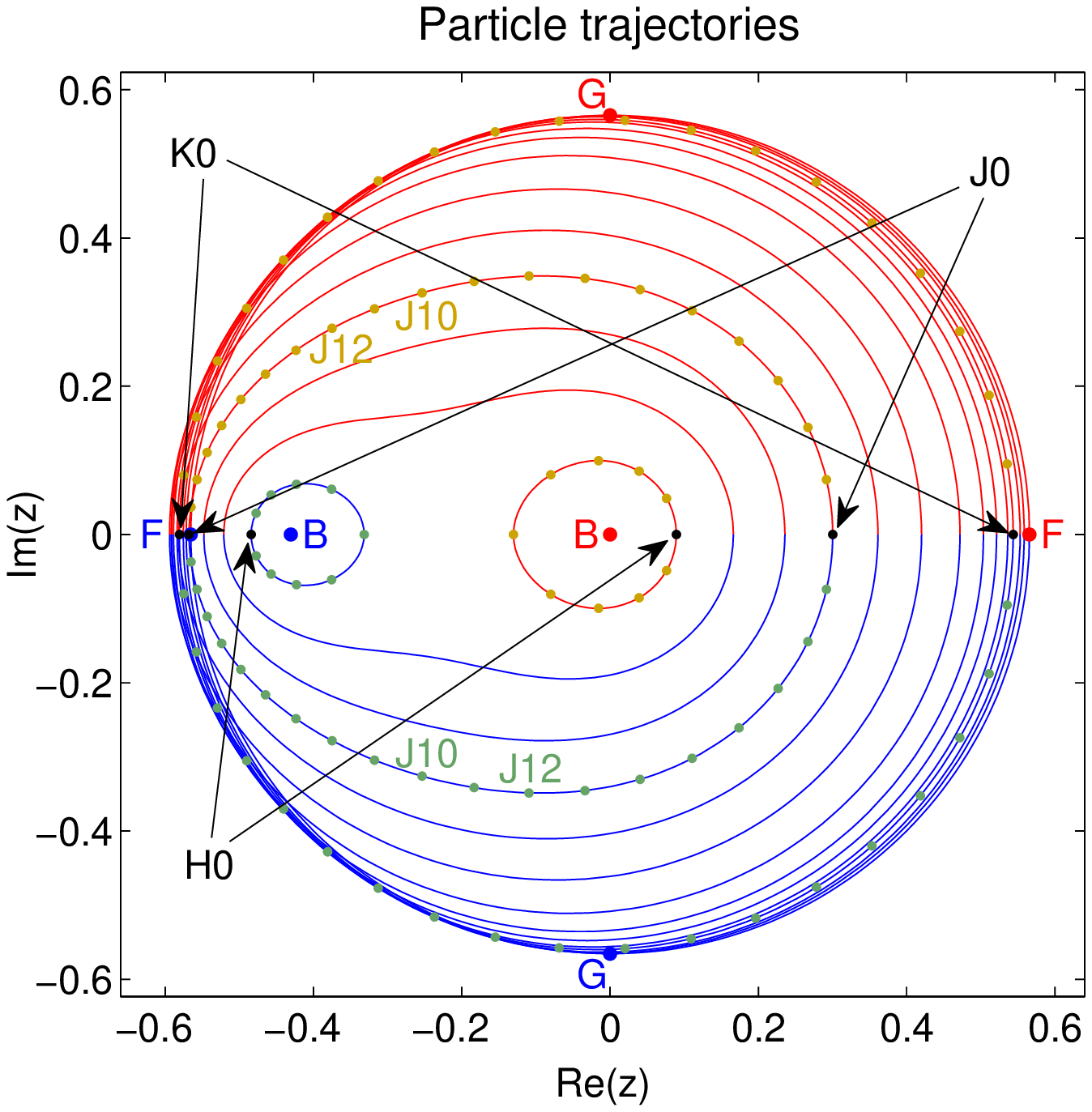}
  \includegraphics[height=2.75in,trim=10 0 0 0,clip]{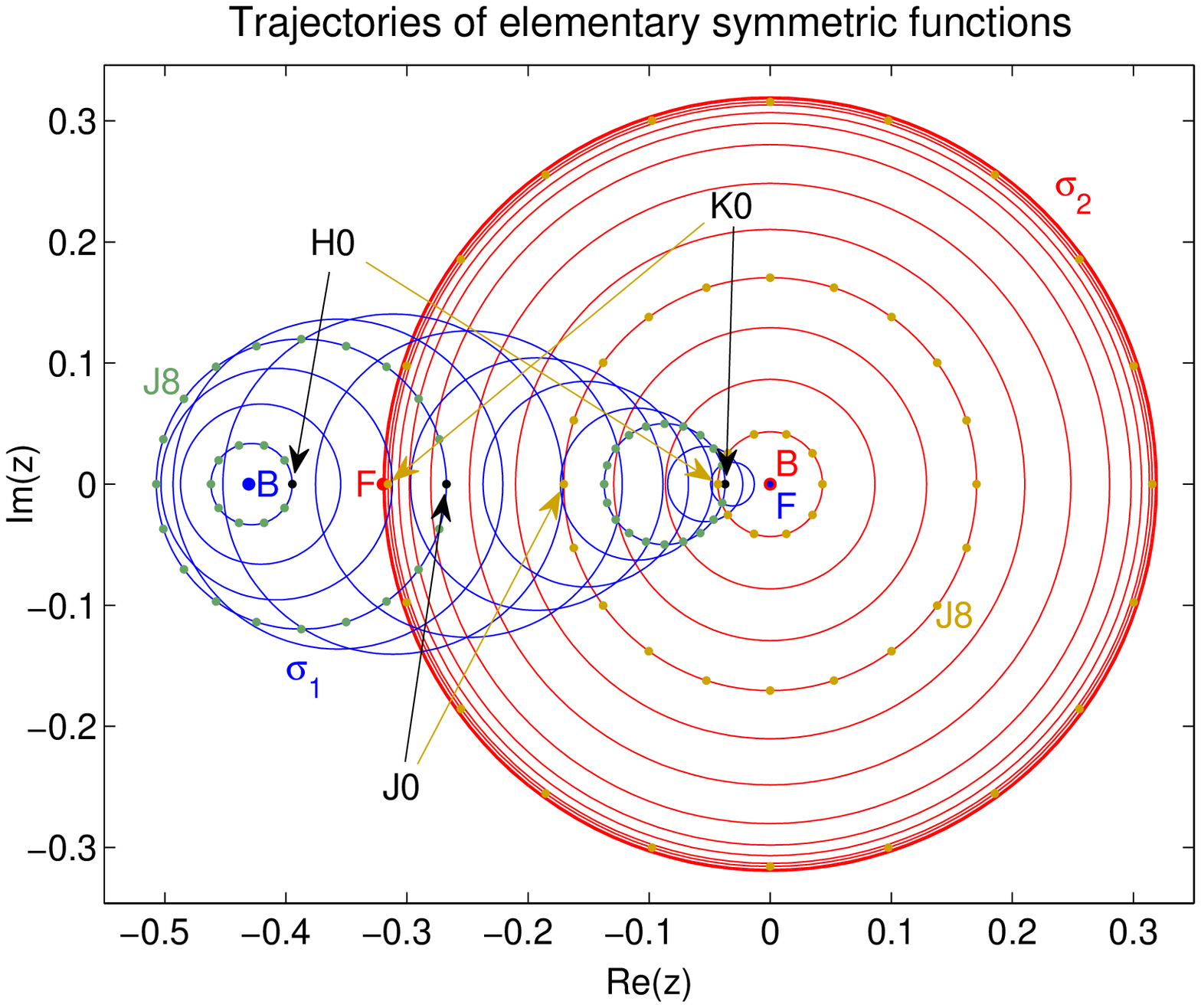}
\end{center}
\caption{\emph{Left:} trajectories of $\beta_1$, $\beta_2$ for the
  solutions in Figures~\ref{fig:evol}--\ref{fig:circles2}.  As we vary
  the bifurcation parameter, the trajectories change from two
  disjoint, counterclockwise loops to one larger orbit in which
  $\beta_1$, $\beta_2$ exchange positions over the course of one
  period.  \emph{Right:} the trajectories of $\sigma_1$, $\sigma_2$
  are exactly circular.}
\label{fig:beta:two:hump}
\end{figure}

Using this approach, we find (numerically) that the solutions on the
path connecting the one-hump stationary solution to the two-hump
traveling wave are $2$-particle solutions.  Moreover, the trajectories
of the first two symmetric functions appear to be of the form
\begin{alignat}{2}
  \label{eqn:sigma1}
  \sigma_1 &= \beta_1+\beta_2 & &= -A + Be^{i\omega t}, \\
  \label{eqn:sigma2}
  \sigma_2 &= \beta_1\beta_2 & &= -Ce^{i\omega t},
\end{alignat}
where $A$, $B$, $C$, $\omega$ are positive constants; see
Figure~\ref{fig:beta:two:hump}.  We now prove this rigorously.

\begin{theorem}
  There is a four-parameter family of time-periodic, two-particle
  solutions of the form
\begin{equation}
  u(x,t) = \alpha_0 + u_{\beta_1(t)}(x) + u_{\beta_2(t)}(x),
\end{equation}
where $\beta_1(t)$ and $\beta_2(t)$ are the roots of the equation
\begin{equation}
  z^2 - \sigma_1(t) z + \sigma_2(t) = 0
\end{equation}
and
\begin{align}
  \sigma_1(t)&=[-A + Be^{i\omega (t-t_0)}]e^{-i\theta}, \qquad
  \sigma_2(t)=[-Ce^{i\omega (t-t_0)}]e^{-2i\theta}, \\[5pt]
  \label{eqn:A:exact}
  A &= \frac{(3-\alpha_0)
    \sqrt{\big[(3-\alpha_0)-(7-\alpha_0)C^2\big]\big[
      (1-\alpha_0)-(5-\alpha_0)C^2\big]}}{
    (3-\alpha_0)^2-(5-\alpha_0)^2C^2}, \\
  \label{eqn:B:exact}
  B &= \frac{5-\alpha_0}{3-\alpha_0} AC, \\[3pt]
  \label{eqn:omega:exact}
  \omega &= \frac{(3-\alpha_0)^2 - (5-\alpha_0)^2 C^2}{
    (3-\alpha_0) - (5-\alpha_0) C^2}.
\end{align}
The four parameters are the mean $\alpha_0<1$, two phases
$\theta,t_0\in\mathbb{R}$, and a real number $C$ ranging from
$C=0$ (at the one-hump stationary solution) to
$C=\sqrt{\frac{1-\alpha_0}{5-\alpha_0}}$ (at the two-hump
traveling wave).
\end{theorem}

\begin{proof}
  It suffices to consider the case that $\theta=0$ and $t_0=0$ as the
  general case follows immediately.  If we try to substitute
  $\beta_{1,2}=
  \frac{\sigma_1}{2}\pm\frac{1}{2}\sqrt{\sigma_1^2-4\sigma_2}$ into
  the system
\begin{align}
  \dot{\beta}_1 = \frac{-2i\beta_1^2}{\beta_1-\beta_2} +
  \frac{2i\beta_1^2}{\beta_1-\bar{\beta}_1^{-1}} +
  \frac{2i\beta_1^2}{\beta_1-\bar{\beta}_2^{-1}} +
  i(3-\alpha_0)\beta_1 \\[5pt]
  \dot{\beta}_2 = \frac{-2i\beta_2^2}{\beta_2-\beta_1} +
  \frac{2i\beta_2^2}{\beta_2-\bar{\beta}_1^{-1}} +
  \frac{2i\beta_2^2}{\beta_2-\bar{\beta}_2^{-1}} +
  i(3-\alpha_0)\beta_2
\end{align}
and solve for $A$, $B$ and $\omega$ in terms of $C$ and $\alpha_0$, the
algebra becomes unmanageable.  However, we can re-write this system in
terms of $\sigma_1$ and $\sigma_2$ to obtain
\begin{align*}
  \dot{\sigma}_1 &= -2i\left\{
    \frac{\beta_1(\beta_1\bar{\beta}_1)}{1-\beta_1\bar{\beta}_1} +
    \frac{\beta_1(\beta_1\bar{\beta}_2)}{1-\beta_1\bar{\beta}_2} +
    \frac{\beta_2(\beta_2\bar{\beta}_1)}{1-\beta_2\bar{\beta}_1} +
    \frac{\beta_2(\beta_2\bar{\beta}_2)}{1-\beta_2\bar{\beta}_2}
  \right\} + i(1-\alpha_0)\sigma_1, \\
  \dot{\sigma}_2 &= -2i\left\{
    \frac{\beta_1\bar{\beta}_1}{1-\beta_1\bar{\beta}_1} +
    \frac{\beta_1\bar{\beta}_2}{1-\beta_1\bar{\beta}_2} +
    \frac{\beta_2\bar{\beta}_1}{1-\beta_2\bar{\beta}_1} +
    \frac{\beta_2\bar{\beta}_2}{1-\beta_2\bar{\beta}_2}
  \right\}\sigma_2 + 2i(2-\alpha_0)\sigma_2.
\end{align*}
The expressions inside braces remain invariant if we interchange
$\beta_1$ and $\beta_2$; hence, they may be written as rational
functions of $\sigma_1$, $\sigma_2$, $\bar{\sigma}_1$,
$\bar{\sigma}_2$.  Explicitly, we have
\begin{align}
\label{eqn:sigma12:dot}
  \dot{\sigma}_1 &= -2i\frac{P_1}{Q} + i(1-\alpha_0)\sigma_1, \qquad\quad
  \dot{\sigma}_2 = -2i\frac{P_2}{Q}\sigma_2 + 2i(2-\alpha_0)\sigma_2,
  \\[3pt]
\notag
  P_1 &= \sigma_1^2\bar{\sigma}_1
  - 2\bar{\sigma}_1\sigma_2
  - 2\sigma_1^3\bar{\sigma}_2
  + 6\sigma_1|\sigma_2|^2
  - \sigma_1\bar{\sigma}_1^2\sigma_2
  + 2\sigma_1^2\bar{\sigma}_1|\sigma_2|^2
  - 2\bar\sigma_1\sigma_2^2\bar\sigma_2
  - 2\sigma_1|\sigma_2|^4, \\[3pt]
\notag
  P_2 &= |\sigma_1|^2\big(1+3|\sigma_2|^2\big)
  + 4|\sigma_2|^2\big(1-|\sigma_2|^2\big) 
  - 2\big(\sigma_1^2\bar{\sigma}_2 + \bar{\sigma}_1^2\sigma_2\big), \\[3pt]
\notag
  Q &= \big(1-|\sigma_2|^2\big)^2
  - |\sigma_1|^2\big(1+|\sigma_2|^2\big)
  + \big(\sigma_1^2\bar{\sigma}_2 + \bar{\sigma}_1^2\sigma_2\big).
\end{align}
Since $Q$ is a product of non-zero terms of the form
$(1-\beta_i\bar{\beta_j})$, it is never zero.  If we assume
$\sigma_1=-A+Be^{i\omega t}$, $\sigma_2=-Ce^{i\omega t}$, and
$C\ne0$, we find that (\ref{eqn:sigma12:dot}) holds as long as
\begin{align}
  \label{eqn:sigma:dot:reduced1}
  \Big[-2P_1 + (1-\alpha_0)\sigma_1Q\Big] +
  \frac{B}{C}\Big[
    -2P_2\sigma_2 + (4-2\alpha_0)\sigma_2Q\Big] &= 0, \\[3pt]
  \label{eqn:sigma:dot:reduced2}
  -2P_2 + (4-2\alpha_0-\omega)Q &= 0.
\end{align}
We eliminated $\omega$ in (\ref{eqn:sigma:dot:reduced1}) using
$\dot\sigma_1=i\omega Be^{i\omega t}=-\frac{B}{C}\dot\sigma_2$.  Next,
we collect terms containing like powers of $e^{i\omega t}$ and set
them each to zero.  This yields 7 polynomial equations in the
variables $A$, $B$, $C$, $\alpha_0$ and $\omega$; however, several of
them are redundant due to relationships such as $Q^\e{-1}=Q^\e1$ in
the decomposition $Q=Q^\e{-1}e^{-i\omega t}+Q^\e0+Q^\e{1}e^{i\omega
  t}$.  Equation (\ref{eqn:sigma:dot:reduced1}) yields 4 such
equations; two of them are satisfied if we choose $B$ as in
(\ref{eqn:B:exact}) while the remaining two are satisfied if we also
choose $A$ as in (\ref{eqn:A:exact}).  With these choices, all three
equations associated with (\ref{eqn:sigma:dot:reduced2}) are satisfied
provided $\omega$ satisfies (\ref{eqn:omega:exact}).  The special
cases \{$C=0$, $A=\sqrt{\frac{1-\alpha_0}{3-\alpha_0}}$, $B=0$\} and
\{$C=\sqrt{\frac{1-\alpha_0}{5-\alpha_0}}$, $A=0$, $B=0$\} are seen to
correspond to the one-hump stationary solution and two-hump traveling
wave, respectively, as discussed in Section~\ref{stationary}.
\end{proof}

We have verified that the curve connecting B to F in the bifurcation
diagram of Figure~\ref{fig:bifur1} is recovered if we set
$\alpha_0=.544375$ and plot $2(-A+B)$ versus $T=2\pi/\omega$ using the
above formulas for $A$, $B$ and $\omega$ with $C$ ranging from 0 to
$\sqrt{\frac{1-\alpha_0}{5-\alpha_0}}$.

\section{Conclusion}\label{conclusion}

We have presented a general method for finding continua of
time-periodic solutions for nonlinear systems of partial differential
equations.  We have used our method to study global paths of
non-trivial time-periodic solutions connecting stationary and
traveling waves of the Benjamin-Ono equation.
In spite of the non-linearity and non-locality of the Benjamin-Ono
equation, these non-trivial solutions can be interpreted as distorted
superpositions of the stationary or traveling waves at each end of the
path.  Our numerical method is accurate enough that we are able to use
data fitting techniques to recognize the analytical form of the
solutions. In particular, the Fourier coefficients $c_k(t)$ of these
solutions follow Ptolemaic orbits of the form
(\ref{eqn:ck:form}).  This led us to reformulate the equations
governing meromorphic pole dynamics to reveal an exact formula for the
solutions on the four-parameter path connecting the one-hump
stationary solution to the two-hump traveling wave.

In the future, we plan to apply this method to more complicated
systems arising in fluid dynamics, namely the vortex sheet and water
wave problems.  This will allow for comparison with prior numerical
and analytical results \cite{HLS2}, \cite{tolandPlotnikov},
\cite{tolandPlotnikovIooss}.  Additionally, as the Benjamin-Ono
equation is meant as a model for internal waves in a deep, stratified
fluid, it will be of interest to compare time-periodic vortex sheets
and water waves with time-periodic solutions of Benjamin-Ono.

\bibliographystyle{alpha}
\bibliography{refs}

\end{document}